\documentclass[12pt]{article}
\usepackage{cite}
\usepackage{amsmath}
\usepackage{amssymb}
\usepackage{bm}
\usepackage{graphicx}
\numberwithin{equation}{section}

\newcounter{aff}

\begin{document}
\begin{titlepage}
\begin{flushright}
{\footnotesize OCU-PHYS 602}
\end{flushright}
\begin{center}
{\Large\bf
Finiteness and Uniqueness of Duality Cascades\\[6pt]
in Three Dimensions for Affine Quivers
}\\
\bigskip\bigskip
{\large
Sanefumi Moriyama\footnote{\tt moriyama@omu.ac.jp} and
Kichinosuke Otozawa\footnote{\tt se24105d@st.omu.ac.jp}
}\\
\bigskip
{\it Department of Physics, Graduate School of Science, Osaka Metropolitan University,}\\
{\it Sumiyoshi-ku, Osaka, Japan 558-8585}\\[3pt]
\end{center}

\begin{abstract}
For three-dimensional circular-quiver supersymmetric Chern-Simons theories, the questions, whether duality cascades always terminate and whether the endpoint is unique, were rephrased into the question whether a polytope defined in the parameter space of relative ranks for duality cascades is a parallelotope, filling the space by discrete translations.
By regarding circular quivers as affine Dynkin diagrams, we generalize the arguments into other affine quivers.
We find that, after rewriting properties into the group-theoretical language, most arguments work in the generalizations.
Especially, we find that, instead of the original relation to parallelotopes, the corresponding polytope still fills the parameter space but with some gaps.
This indicates that, under certain restrictions, duality cascades still terminate uniquely.
\end{abstract}
\end{titlepage}

\tableofcontents

\section{Introduction}

Duality cascades provide interesting scenarios for theoretical physics.
If a four-dimensional supersymmetric gauge theory is strongly-coupled, sometimes we can dualize it to a weakly-coupled theory.
Nevertheless, under the renormalization group flow, the weakly-coupled theory can become strongly-coupled again.
Subsequent applications of dualities are known as duality cascades \cite{KS}.
It would be interesting if the ranks of gauge groups of the standard model are obtained through duality cascades, as suggested in \cite{St}.

In three dimensions the situation is somewhat different.
Here the duality is an IR duality, which means that two theories with different looks in UV flow to the same IR physics.
Although duality cascades in three dimensions do not necessarily interchange strong/weak theories and evolve alternatingly under renormalization group flows, the UV description with lower ranks is more concise without extra degrees of freedom and should be better.

The duality has an interesting interpretation in brane configurations.
In \cite{HW} a brane configuration of D3-branes with two kinds of 5-branes placed perpendicularly and tilted relatively to preserve supersymmetries was considered.
Then, the three-dimensional supersymmetric gauge theory is realized as the low-energy effective theory on the D3-branes and the duality is realized as an exchange of two neighboring 5-branes known as the Hanany-Witten (HW) transition.
Exactly the same brane configuration placed on a circle revives in the study of the worldvolume theory of M2-branes, the Aharony-Bergman-Jafferis-Maldacena (ABJM) theory \cite{ABJM,HLLLP2,ABJ}.
On the circle we can apply the brane transitions subsequently, which leads to duality cascades \cite{AHHO,EK,HK}.

The ABJM theory is the three-dimensional ${\cal N}=6$ supersymmetric Chern-Simons theory which has gauge group $\text{U}(N)_{-k}\times\text{U}(N+M)_{k}$ (with subscripts denoting the Chern-Simons levels) and two pairs of bifundamental matters.
By applying the localization theorem, it was found that the partition function originally defined by the infinite-dimensional path integral reduces to a finite-dimensional matrix integration.
After an intensive study of the matrix model, finally it was found that, aside from the behavior of $N^{\frac{3}{2}}$ \cite{DMP1,DMP2}, all the perturbative corrections are summed up to the Airy function \cite{FHM,MP} and the non-perturbative corrections are given by the free energy of topological strings on local ${\mathbb P}^1\times{\mathbb P}^1$ \cite{MP,HMO2,HMO3,HMMO}.

The studies of the partition function for the case of equal ranks $M=0$ were based on the 't Hooft expansion \cite{DMP1,DMP2,FHM}, the WKB expansion \cite{MP,CM} and the exact values \cite{PY,HMO1}.
Most of these studies rely heavily on the so-called Fermi gas formalism \cite{MP}, where the grand canonical partition function is regarded as that of one-dimensional non-interacting fermions.
For rank differences, there are two methods.
One of them is based on the open-string viewpoint, where the corrections from the equal-rank case resemble the Wilson loop insertions \cite{MM,HO,MNN}, and the other is based on the closed-string viewpoint, which corrects coefficients of the spectral operator in the Fredholm determinant \cite{MS2,MN5,KM,KMZ2}.
The convergence condition was, however, not very clear in both methods and conservatively the apparent convergence condition $0\le M\le k/2$ had been adopted \cite{HMO2} for a while.

There are many generalizations of the ABJM theory represented by a circular quiver with two nodes in the quiver diagram.
First, it is natural to generalize the studies to circular quivers with more nodes.
Second, by regarding the circular quivers as the affine Dynkin diagrams of type $\widehat A$, it is natural to generalize the studies to quivers identical to other affine Dynkin diagrams such as those of types $\widehat D$ and $\widehat E$ \cite{GHN}.
Though the supersymmetries are broken to ${\cal N}=3$ generally in these generalizations, they are enhanced to ${\cal N}=4$ when there are only two kinds of 5-branes in the circular quivers \cite{IK}.
The simply-laced affine quiver diagrams of $\widehat A\widehat D\widehat E$ are special in the sense that the beta functions for unitary groups vanish identically for the equal-rank cases.
Other non-simply-laced affine quiver diagrams work as well after introducing orthogonal or symplectic groups.

All of these theories were studied intensively with different levels of achievement.
Especially, for the ${\cal N}=4$ enhanced circular quivers of type $\widehat A$, partition functions were studied extensively \cite{MN1,MN3,HHO}.
Partition functions in the grand canonical ensemble, after normalized by the lowest order, are rewritten into the Fredholm determinant of spectral operators.
Interestingly, the spectral operators sometimes enjoy symmetries of exceptional Weyl groups \cite{MP,MN3,MNN,MNY,KMN,MS,MY}, which suggests a relation to $q$-Painlev\'e equations.
Indeed, it was found that the grand partition function of the ABJM theory satisfies the $q$-Painlev\'e equation $q\text{P}_{\text{III}_3}$ \cite{BGT}, while that of the generalized one with the numbers of 5-branes doubled satisfies $q\text{P}_\text{VI}$ \cite{BGKNT,MN6}.  

Since the brane configuration is placed on a circle, there seem to be infinitely many non-commuting applications of the HW transitions in duality cascades and in general it is unclear whether duality cascades always terminate in finite steps and whether the final destination of duality cascades is uniquely determined from an initial configuration regardless of various intermediate processes.
These questions were answered positively for general ${\cal N}=3$ circular quivers in \cite{FMS}.
To avoid confusions we first need to fix the working hypothesis of duality cascades \cite{FMMN} as follows.
Preliminarily we fix one of the lowest ranks to be the reference \cite{KuMo} and repeat the following process as many times as possible.
(1) We apply the HW transitions arbitrarily without crossing the reference.
(2) We change references by cyclic rotations if we encounter lower ranks compared with the original reference.
By reordering the 5-branes in the standard order, we can regard duality cascades as transformations in the space of ranks.
Then, the fundamental domain of duality cascades is defined to be the parameter domain in the space of relative ranks where no more duality cascades are applicable.
In terms of it, the above questions on the finiteness and the uniqueness of duality cascades are rephrased geometrically into whether the fundamental domain forms a parallelotope, a polytope tiling the entire parameter space by discrete translations.
It was proved \cite{FMS} that the fundamental domain is a parallelotope, by first proposing three descriptions including the description by zonotopes and confirming the space-filling criterion for zonotopes \cite{Sh,M}.
Hence, the above questions on the finiteness and the uniqueness are answered positively.
Interestingly, from one of the descriptions, the fundamental domain was also observed to be a permutohedron.

For some special cases enjoying the symmetries of exceptional Weyl groups, the fundamental domains are the affine Weyl chambers \cite{FMMN}.
For these cases, some of the Weyl reflections correspond to the HW transitions \cite{KuMo}, which implies a bigger picture to unify studies of M2-branes from the symmetry viewpoints.
These viewpoints also help to clarify various computations of partition functions performed so far.
Previously, observations that the grand partition functions satisfy the $q$-Painlev\'e equations \cite{BGT,MN6} were confirmed only in the original apparent convergence domain.
With the geometric viewpoints, we can further extend the confirmation \cite{BGT,MN6} into the whole parameter space \cite{MN7}.

Compared with the extensive studies for the circular quivers of type $\widehat A$, the situation for other quivers is premature.
For the quivers of type $\widehat D$, the Fermi gas formalism was proposed in \cite{ADF,MN4}.
In \cite{MN4,KN} it was further found that, with a special choice of levels $(-k,0,0,\cdots,0,k)$, we can generalize the WKB analysis in \cite{MP} and analyze membrane instantons order by order.

Hence, as the next step of the progress, it is crucial to clarify the following questions.
\begin{itemize}
\item
How do duality cascades apply to other affine quivers such as those of types $\widehat D$ and $\widehat E$?
Especially, for these quivers, do duality cascades always terminate in finite steps as well and is the final destination uniquely determined by an initial configuration regardless of various processes of duality cascades?
\item
Regarding the analysis of the fundamental domains, technically, how are the three descriptions of the fundamental domain for the $\widehat A$ quivers generalized to those for the $\widehat D$ and $\widehat E$ quivers?
Especially, what is the relation to permutohedrons for the $\widehat A$ quivers \cite{FMS} generalized for the $\widehat D\widehat E$ quivers?
\item
Why is the choice of levels $(-k,0,0,\cdots,0,k)$ special for the $\widehat D$ quivers in solving the matrix models?
Can we understand the choice geometrically?
\item
How about non-simply-laced quivers?
Can we discuss duality cascades for them as well?
\end{itemize}

In this paper, we shall study duality cascades for other affine quivers such as the $\widehat D$ and $\widehat E$ quivers or the non-simply-laced quivers.
The key point in generalizations is the group theory. 
We first revisit duality cascades for the $\widehat A$ quivers by fully adopting the group-theoretical language.
Namely, we reinterpret the choice of the reference rank as the choice of the affine node and the application of the cyclic rotations as the application of the outer automorphism for the affine Dynkin diagram.
Thus, the generalizations of duality cascades become clearer.
Then, we encounter the same questions, whether duality cascades always terminate in finite steps and whether the endpoint is unique.
We answer these questions by the same geometric methods.

The main difference between the $\widehat A$ quivers and the other quivers is the non-trivial marks, which are the Dynkin labels for the affine root, the expansion coefficients with respect to all the finite simple roots. 
For the $\widehat A$ quivers all of the marks are trivially one and all the nodes can potentially be the affine node.
Therefore, duality cascades work on all the nodes.
This is not the case for the other quivers.
Besides, the non-trivial marks introduce gaps for the fundamental domain when tiling the whole parameter space.
Indeed we find that, these non-trivial marks appear in some discrete translations of duality cascades, while they are absent in the distance between the corresponding opposite parallel facets.
These discrepancies are responsible for the gaps.

After clarifying the behavior in duality cascades, we also comment on possible directions for studies of partition functions from this geometric viewpoint.
We observe that our generalization into generic levels introduces a polytope described clearly in the group-theoretical language, which tiles the space with gaps, and the choice of levels $(-k,0,0,0,k)$ for the $\widehat D_4$ quiver studied in \cite{MN4,KN} fills the gaps.
This viewpoint should enable further studies of partition functions in duality cascades.

We also study the fundamental domain for the $\widehat B_3$ and $\widehat C_3$ quivers.
It is interesting to note that, although the finite Weyl groups for algebras $B$ and $C$ are identical, the affine Weyl groups are different.
This implies that, although the $\widehat B$ and $\widehat C$ quivers share the same fundamental domains as zonotopes, the discrete translations of duality cascades may work differently.
Indeed, we find that, for the two identical fundamental domains, the two distinct discrete translations lead to different patterns of tiling.

This paper is organized as follows.
In the next section, we start with a review of duality cascades for the supersymmetric Chern-Simons theories of the $\widehat A$ quivers.
Instead of the original language of brane configurations, we rewrite the working hypothesis of duality cascades fully in terms of the group-theoretical language, which is more applicable to generalizations to those of the $\widehat D$ and $\widehat E$ quivers.
Then, we embark on studies of duality cascades for the $\widehat D$ and $\widehat E$ quivers respectively in sections \ref{D} and \ref{E}.
We also comment on possible implications to partition functions from this geometric viewpoint in section \ref{pf}.
Finally we conclude with some future directions.
The appendix is devoted to the study of duality cascades for the non-simply-laced quivers $\widehat B_3$ and $\widehat C_3$.

\section{$\widehat A$ quivers in group theories}

In this section we shall review duality cascades for the $\widehat A$ quivers \cite{FMMN,FMS} and rewrite them into the group-theoretical language.
As we see below, understanding from the group theory not only helps in applying duality cascades to other affine quivers such as the $\widehat D\widehat E$ quivers, but also refines our understandings on the fundamental domain of duality cascades, an associated polytope defined to answer the questions on the finiteness and the uniqueness raised in the introduction, even for the original $\widehat A$ quivers.

\subsection{HW transitions and duality cascades}

\begin{figure}[t!]
\centering\includegraphics[scale=0.5,angle=-90]{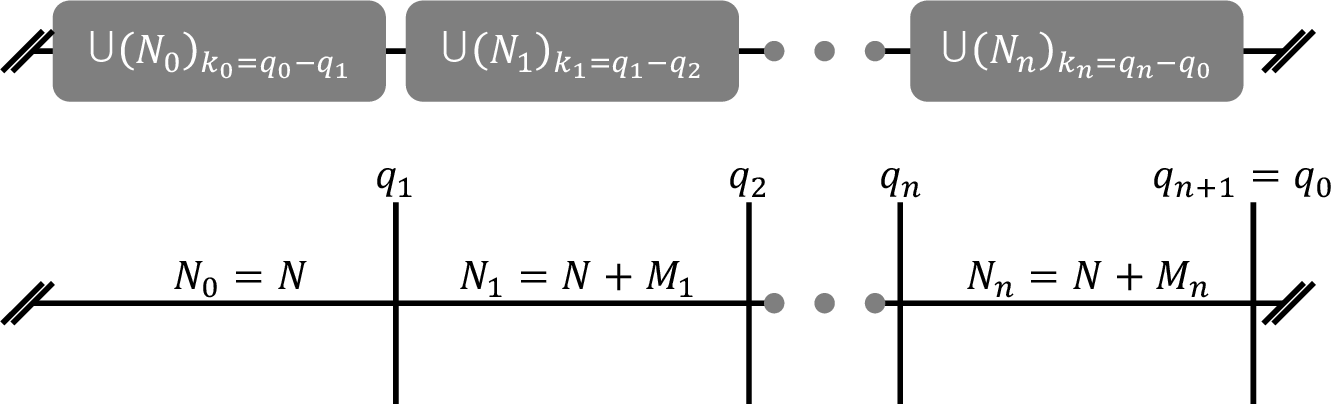}
\caption{A circular quiver diagram (top) and the corresponding circular brane configuration (bottom).
The circular-quiver super Chern-Simons theory with levels and ranks specified is realized as the effective theory for the brane configuration.
In the brane configuration, the vertical lines are various 5-branes of charge $(1,q_i)$ and the horizontal lines are D3-branes compactified in one dimension where the number of D3-branes can be different for each interval.
We choose the lowest rank to be the reference $N_0=N$ and label the other ranks by the differences $N_i=N+M_i$.
Later we often denote the brane configuration by $\langle N_0\stackrel{q_1}{\bullet}N_1\stackrel{q_2}{\bullet}\cdots\stackrel{q_n}{\bullet}N_n\stackrel{q_{n+1}}{\bullet}\rangle$ or $\langle\stackrel{q_1}{\bullet}M_1\stackrel{q_2}{\bullet}\cdots\stackrel{q_n}{\bullet}M_n\stackrel{q_{n+1}}{\bullet}\rangle$ when the overall rank $N$ is irrelevant.}
\label{An}
\end{figure}

In this section, we consider the ${\cal N}=3$ super Chern-Simons theories with gauge group $\text{U}(N_0)_{k_0}\times\text{U}(N_1)_{k_1}\times\cdots\times\text{U}(N_n)_{k_n}$ and pairs of bifundamental matters connecting adjacent group factor cyclically.
The subscripts $k_i$ of $\text{U}(N_i)_{k_i}$ denote the Chern-Simons levels subject to the zero sum condition $\sum_{i=0}^nk_i=0$.
The theories are often characterized by quiver diagrams (see figure \ref{An}).
Since the theories are realized as low-energy effective theories in brane configurations, we start with the explanation of the brane configuration (see figure \ref{An} for the brane configuration and table \ref{branedirection} for the directions various branes extend to).
The brane configuration in IIB string theory consists of D3-branes with 5-branes of various charges placed perpendicularly and tilted relatively to preserve supersymmetries.
The 5-branes are located at different positions in the direction 6 and D3-branes end on these 5-branes.

\begin{table}
\centering
\begin{tabular}{c|ccccc}
branes&0\;1\;2&6&3\;7&4\;8&5\;9\\\hline
D3&$-$$-$$-$&$-$&&&\\
NS5&$-$$-$$-$&&$-\quad$&$-\quad$&$-\quad$\\
D5&$-$$-$$-$&&$\quad -$&$\quad -$&$\quad -$\\
$(p,q)$5&$-$$-$$-$&&$[3,7]_\theta$&$[4,8]_\theta$&$[5,9]_\theta$
\end{tabular}
\caption{The directions which various branes extend to.
$-$ denotes the directions where branes extend and $[i,j]_\theta$ stands for the direction $i$ tilted to $j$ by an angle $\theta$.
Namely, aside from directions $0,1,2$, $(p,q)$5-branes extend to directions $3,4,5$ tilted to $7,8,9$ with a common angle $\theta=\arctan q/p$.}
\label{branedirection}
\end{table}

It is known that the brane configuration enjoys symmetries of exchanging 5-branes, the HW brane transitions \cite{HW}.
The HW transition \cite{HW} is the change of ranks caused by the change of 5-brane orders.
When we switch two adjacent 5-branes, the intermediate rank is changed as
\begin{align}
\cdots K\bullet L\circ M\cdots=\cdots K\circ K-L+M+|k|\bullet M\cdots.
\label{hw}
\end{align}
Here $\bullet$ and $\circ$ denote two 5-branes and the level $k$ is fixed by the determinant of their 5-brane charges.
Especially, when the two 5-branes are of the same kind, $k$ is vanishing, which we also include in the HW transitions.
As already pointed out in \cite{HW}, the HW transitions can be understood as charge conservation.
Namely, the HW transition \eqref{hw} is derived by requiring the conservation of the RR charge $Q_\text{RR}$ for each NS5-brane.
Due to the charge conservation, although there are two paths to change the order of 5-branes from $\stackrel{q_1}{\bullet}\stackrel{q_2}{\bullet}\stackrel{q_3}{\bullet}$ to $\stackrel{q_3}{\bullet}\stackrel{q_2}{\bullet}\stackrel{q_1}{\bullet}$ by the exchanges $\stackrel{q_1}{\bullet}\leftrightarrow\stackrel{q_2}{\bullet}$, $\stackrel{q_1}{\bullet}\leftrightarrow\stackrel{q_3}{\bullet}$, $\stackrel{q_2}{\bullet}\leftrightarrow\stackrel{q_3}{\bullet}$ and $\stackrel{q_2}{\bullet}\leftrightarrow\stackrel{q_3}{\bullet}$, $\stackrel{q_1}{\bullet}\leftrightarrow\stackrel{q_3}{\bullet}$, $\stackrel{q_1}{\bullet}\leftrightarrow\stackrel{q_2}{\bullet}$, the final result does not depend on the paths.
Hereafter we often consider 5-branes with collinear charges $(p,q)=(1,q_i)$ and denote 5-branes by $\stackrel{q_i}{\bullet}$ with labels of the charges.

For the ABJM theory \cite{ABJM} or more general worldvolume theories of M2-branes, we need to place the brane configuration on a circle by compactifying the direction 6.
Then, on the circle, we can continue to perform the HW transitions \eqref{hw} infinitely.
In supersymmetric gauge theories, an application of the HW transition corresponds to a duality transformation, whose series combine into duality cascades explained below.

In \cite{FMMN,FMS} the working hypothesis of duality cascades was proposed as follows.
Preliminarily, we assign one of the lowest ranks as the reference rank temporally.
\begin{itemize}
\item[$1.$] We apply the HW transitions \eqref{hw} arbitrarily without crossing the reference.
\item[$2.$] We change references by cyclic rotations when encountering lower ranks compared with the current reference.
\end{itemize}
We regard the above process as one step of duality cascades and continue as many steps as possible until we encounter no lower ranks compared with the reference.
Since we can apply the HW transitions arbitrarily, we often compare configurations by fixing orders of 5-branes.

In order to generalize to the $\widehat D$ and $\widehat E$ quivers later, let us rewrite the process in the working hypothesis into the group-theoretical language.
We first reinterpret the HW transition \eqref{hw} as the Weyl reflection.
We consider the brane configuration with 5-branes having collinear charges $(1,q_i)$.
The levels fixed from the determinant of 5-brane charges $(1,q_i)$ and $(1,q_{i+1})$ of two adjacent 5-branes, $\stackrel{q_i}{\bullet}$ and $\stackrel{q_{i+1}}{\bullet}$, are then given by $k_i=q_i-q_{i+1}$.
By exchanging the two adjacent 5-branes in
$\cdots\stackrel{q_{i-1}}{\bullet}\cdot\stackrel{q_i}{\bullet}\cdot\stackrel{q_{i+1}}{\bullet}\cdot\stackrel{q_{i+2}}{\bullet}\cdots$ giving rise to levels
\begin{align}
\cdots\times\text{U}(\#)_{k_{i-1}=q_{i-1}-q_i}\times\text{U}(\#)_{k_i=q_i-q_{i+1}}\times\text{U}(\#)_{k_{i+1}=q_{i+1}-q_{i+2}}\times\cdots,
\label{bef}
\end{align}
with ranks omitted, we trivially obtain $\cdots\stackrel{q_{i-1}}{\bullet}\cdot\stackrel{q_{i+1}}{\bullet}\cdot\stackrel{q_i}{\bullet}\cdot\stackrel{q_{i+2}}{\bullet}\cdots$ giving rise to levels 
\begin{align}
\cdots\times\text{U}(\#)_{k_{i-1}+k_i=q_{i-1}-q_{i+1}}\times\text{U}(\#)_{-k_i=q_{i+1}-q_i}\times\text{U}(\#)_{k_{i+1}+k_i=q_i-q_{i+2}}\times\cdots.
\label{aft}
\end{align}
Apparently, the change of levels from $(\cdots,k_{i-1},k_i,k_{i+1},\cdots)$ to $(\cdots,k_{i-1}+k_i,-k_i,k_{i+1}+k_i,\cdots)$ is isomorphic to the Weyl reflection, which reflects weights $w$ with respect to the hyperplane perpendicular to the simple root $\alpha_i$,
\begin{align}
w\mapsto w-2\frac{(w,\alpha_i)}{(\alpha_i,\alpha_i)}\alpha_i,
\label{weylrefl}
\end{align}
and maps simple roots of the $A$ algebra $(\cdots,\alpha_{i-1},\alpha_i,\alpha_{i+1},\cdots)$ to $(\cdots,\alpha_{i-1}+\alpha_i,-\alpha_i,\alpha_{i+1}+\alpha_i,\cdots)$.
The introduction of the ranks $N_i$ in \eqref{hw} does not ruin the algebras of the Weyl reflections and finally we have
\begin{align}
r_i:k_{i-1}\to k_{i-1}+k_i,\;k_i\to-k_i,\;k_{i+1}\to k_{i+1}+k_i,\;N_i\to N_{i-1}-N_i+N_{i+1}+|k_{i}|.
\label{aweyl}
\end{align}
The previous comment below \eqref{hw} that the ranks do not depend on the processes of exchanging 5-branes is understood as the commutation relation of Weyl reflections $r_ir_{i+1}r_i=r_{i+1}r_ir_{i+1}$ or $(r_ir_{i+1})^3=1$ for the neighboring ones.
After reinterpreting the HW transition \eqref{hw} as the Weyl reflection \eqref{aweyl}, we often abuse these two words interchangeably.

In the group-theoretical language, we can rewrite the process of duality cascades as follows.
We first regard the circular quiver as the affine Dynkin diagram of type $\widehat A$.
Although in the $\widehat A$ quiver all of the nodes are equivalent, let us fix one node with the lowest rank as the affine node preliminary, as we have chosen the lowest rank in the circular quiver to be the reference.
\begin{itemize}
\item[$1'.$] We apply finite Weyl reflections arbitrarily (without using the affine one).  
\item[$2'.$] We change affine nodes by applying the outer automorphisms of the affine Weyl group, when encountering lower ranks compared with that of the current affine node.
\end{itemize}
Note that the previous choice of the reference is now rewritten into the choice of the affine node, the previous condition of avoiding crossing the reference in exchanging 5-branes is rewritten into the condition of avoiding the use of the affine Weyl reflection and the previous use of the cyclic rotation in changing references is now the use of the outer automorphism.

\begin{table}[t!]
\begin{eqnarray*}
&\rotatebox[origin=r]{30}{=}\;\langle 5\stackrel{3}{\bullet}3\stackrel{0}{\bullet}9\stackrel{2}{\bullet}\rangle\to\langle 3\stackrel{0}{\bullet}9\stackrel{2}{\bullet}5\stackrel{3}{\bullet}\rangle=\langle 3\stackrel{2}{\bullet}1\stackrel{3}{\bullet}2\stackrel{0}{\bullet}\rangle\to\langle 1\stackrel{3}{\bullet}2\stackrel{0}{\bullet}3\stackrel{2}{\bullet}\rangle\;\rotatebox[origin=l]{-30}{=}&\\
\langle 5\stackrel{0}{\bullet}14\stackrel{2}{\bullet}11\stackrel{3}{\bullet}\rangle\hspace{-20mm}&&\hspace{-20mm}\langle 1\stackrel{0}{\bullet}5\stackrel{2}{\bullet}4\stackrel{3}{\bullet}\rangle\\
&\rotatebox[origin=r]{-30}{=}\;\langle 5\stackrel{2}{\bullet}4\stackrel{0}{\bullet}11\stackrel{3}{\bullet}\rangle\to\langle 4\stackrel{0}{\bullet}11\stackrel{3}{\bullet}5\stackrel{2}{\bullet}\rangle=\langle 4\stackrel{3}{\bullet}1\stackrel{0}{\bullet}5\stackrel{2}{\bullet}\rangle\;\rotatebox[origin=l]{30}{$\to$}&
\end{eqnarray*}
\caption{Two processes of duality cascades starting from the brane configuration $\langle 5\stackrel{0}{\bullet}14\stackrel{2}{\bullet}11\stackrel{3}{\bullet}\rangle$.
We denote the exchange of 5-branes with the reference fixed (or the finite Weyl reflections) by ``$=$'' and the cyclic rotation (or the outer automorphism) by ``$\to$''.}
\label{ex}
\end{table}

Let us take the brane configuration with three 5-branes as an example, $\langle 5\stackrel{0}{\bullet}14\stackrel{2}{\bullet}11\stackrel{3}{\bullet}\rangle$, and explain the above working hypothesis of duality cascades (see table \ref{ex}).
Here we label 5-branes of charges $(1,q_i)$ by $\stackrel{q_i}{\bullet}$ and denote the reference (or the affine node) as the endpoints in the bracket.
Physically, this brane configuration corresponds to the gauge group $\text{U}(5)_3\times\text{U}(14)_{-2}\times\text{U}(11)_{-1}$ with pairs of bifundamental matters denoted in figure \ref{An}.
We can bring the rightmost 5-brane $\stackrel{3}{\bullet}$ to the leftmost using HW transitions \eqref{hw} (or equivalently apply Weyl reflections $r_1r_2$ in \eqref{aweyl} subsequently) and obtain the configuration $\langle 5\stackrel{3}{\bullet}3\stackrel{0}{\bullet}9\stackrel{2}{\bullet}\rangle$.
Since we encounter a lower rank than the reference, let us apply the cyclic rotation (equivalent to the outer automorphism of the affine Dynkin diagram) to find the configuration $\langle 3\stackrel{0}{\bullet}9\stackrel{2}{\bullet}5\stackrel{3}{\bullet}\rangle$.
Furthermore, we can continue more steps in the working hypothesis and reach the configuration $\langle 1\stackrel{3}{\bullet}2\stackrel{0}{\bullet}3\stackrel{2}{\bullet}\rangle$.
Note that now we cannot find lower ranks compared with the reference rank for this configuration by applying any combinations of HW transitions without crossing the reference (or any combinations of finite Weyl reflections) and hence duality cascades terminate here.
For comparison, we bring 5-branes into the original order of $\langle\stackrel{0}{\bullet}\stackrel{2}{\bullet}\stackrel{3}{\bullet}\rangle$ in the final expression.
Alternatively, we can follow other processes in duality cascades such as that given in table \ref{ex}.
Even though the processes are different, we end up with exactly the same configuration.

It is then natural to ask (F) whether duality cascades always terminate in finite steps and (U) whether the final configuration of duality cascades is always uniquely determined depending only on an initial one regardless of various processes.
These questions were answered positively in \cite{FMS} geometrically, as we recapitulate in the next subsection.

\subsection{Finiteness and uniqueness of duality cascades}\label{finuni}
 
To discuss the above questions on the finiteness and the uniqueness of duality cascades geometrically, let us define the fundamental domain of duality cascades as the parameter domain of relative ranks where no more duality cascades are applicable \cite{FMMN,FMS}.
After understanding from the charge conservation $Q_\text{RR}$ that duality cascades are realized as discrete translations in the parameter space of relative ranks, the above questions on the finiteness and the uniqueness of duality cascades are rephrased as follows.
(F) Starting from an arbitrary point in the parameter space of relative ranks specifying a configuration, does the point always reduce to the fundamental domain by the discrete translations of duality cascades?
Conversely, does the fundamental domain cover the whole parameter space of relative ranks by its infinite copies obtained by the discrete translations?
(U) Is the final destination of the discrete translations uniquely determined only by an initial point regardless of various processes?
Conversely, are the infinite copies of the fundamental domain disjoint, sharing at most the boundaries?
Since a polytope which tiles the entire space by discrete translations mutually exclusively and collectively exhaustively is called a parallelotope, the above two questions can be summarized as whether the fundamental domain forms a parallelotope.

In \cite{FMS} three equivalent descriptions of the fundamental domain were proposed.
Let us explain the three descriptions for the example of the brane configuration
\begin{align}
\langle N\stackrel{q_1}{\bullet}N+M_1\stackrel{q_2}{\bullet} N+M_2\stackrel{q_3}{\bullet}\rangle,
\label{three5}
\end{align}
or equivalently for the gauge theory with gauge group $\text{U}(N)_{-q_{13}}\times\text{U}(N+M_1)_{q_{12}}\times\text{U}(N+M_2)_{q_{23}}$ and bifundamental matters.
Here we have introduced the notation $q_{ij}=q_i-q_j$.

\begin{table}[t!]
\centering
\begin{tabular}{l|r}
brane configurations&Weyl reflections\\\hline
$\langle\stackrel{q_1}{\bullet}M_1\stackrel{q_2}{\bullet}M_2\stackrel{q_3}{\bullet}\rangle$&$1$\\
$\stackrel{1\leftrightarrow 2}{=}\langle\stackrel{q_2}{\bullet}-M_1+M_2+|q_{12}|\stackrel{q_1}{\bullet}M_2\stackrel{q_3}{\bullet}\rangle$&$r_1$\\
$\stackrel{1\leftrightarrow 3}{=}\langle\stackrel{q_2}{\bullet}-M_1+M_2+|q_{12}|\stackrel{q_3}{\bullet}-M_1+|q_{12}|+|q_{13}|\stackrel{q_1}{\bullet}\rangle$&$r_2r_1$\\
$\stackrel{2\leftrightarrow 3}{=}\langle\stackrel{q_3}{\bullet}-M_2+|q_{13}|+|q_{23}|\stackrel{q_2}{\bullet}-M_1+|q_{12}|+|q_{13}|\stackrel{q_1}{\bullet}\rangle$&$r_1r_2r_1$\\
$\stackrel{2\leftrightarrow 1}{=}\langle\stackrel{q_3}{\bullet}-M_2+|q_{13}|+|q_{23}|\stackrel{q_1}{\bullet}M_1-M_2+|q_{23}|\stackrel{q_2}{\bullet}\rangle$&$r_2r_1r_2r_1$\\
$\stackrel{3\leftrightarrow 1}{=}\langle\stackrel{q_1}{\bullet}M_1\stackrel{q_3}{\bullet}M_1-M_2+|q_{23}|\stackrel{q_2}{\bullet}\rangle$&$r_1r_2r_1r_2r_1$
\end{tabular}
\caption{Various expressions of the brane configuration $\langle\stackrel{q_1}{\bullet}M_1\stackrel{q_2}{\bullet}M_2\stackrel{q_3}{\bullet}\rangle$ obtained by the HW transitions \eqref{hw} or the Weyl reflections \eqref{aweyl}.
We abbreviate the exchange $\stackrel{i}{\bullet}\leftrightarrow\stackrel{j}{\bullet}$ as $i\leftrightarrow j$ and attach it to the Weyl reflection ``$=$''.}
\label{a2h}
\end{table}

The first one (the ${\cal H}$ description) utilizes the original definition of duality cascades.
Namely, we consider the conditions that no lower ranks appear through the HW transitions (see table \ref{a2h}).
We change the order of 5-branes arbitrarily using the HW transitions \eqref{hw} (or apply the Weyl reflections \eqref{aweyl}) for the brane configuration \eqref{three5} and require that no ranks lower than the reference (or the affine node) appear.
Since the overall rank $N$ is not affected by these exchanges, we subtract it from the beginning and denote \eqref{three5} by $\langle\stackrel{q_1}{\bullet}M_1\stackrel{q_2}{\bullet}M_2\stackrel{q_3}{\bullet}\rangle$.
From table \ref{a2h} we can read off the inequalities
\begin{align}
&M_1\ge 0,\quad M_2\ge 0,\quad
-M_1+M_2+|q_{12}|\ge 0,\nonumber\\
&{-M_1}+|q_{12}|+|q_{13}|\ge 0,\quad
-M_2+|q_{13}|+|q_{23}|\ge 0,\quad
M_1-M_2+|q_{23}|\ge 0.
\label{a2ineq}
\end{align}
The fundamental domain of duality cascades is defined as the domain in the $(M_1,M_2)$ space satisfying these inequalities, each of which restricts to half of the space.

\begin{figure}[t!]
\centering\includegraphics[scale=0.5,angle=-90]{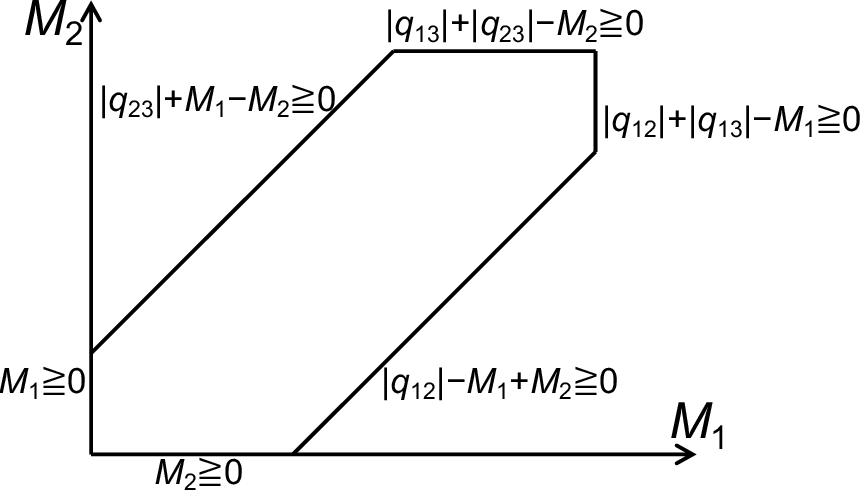}\includegraphics[scale=0.5,angle=-90]{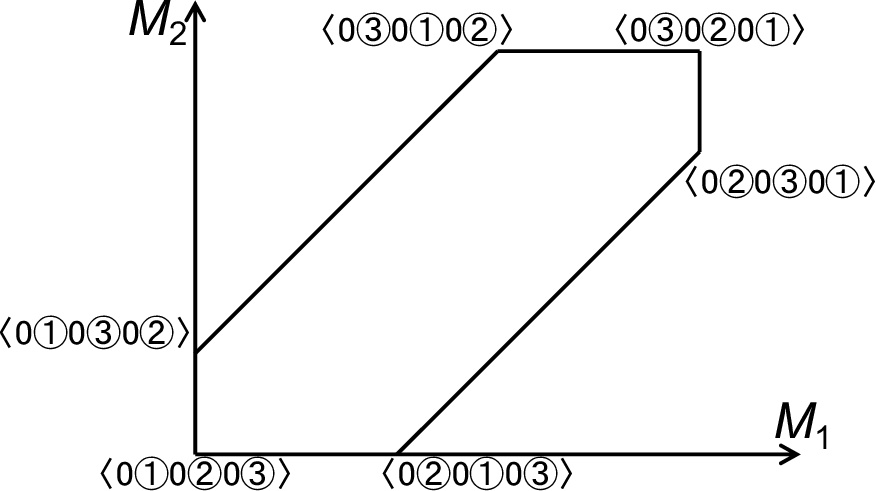}\\[6pt]
(${\cal H}$)\hspace{6cm}(${\cal V}$)\\[6pt]
\rotatebox[origin=c]{-45}{\Huge =}\quad\includegraphics[scale=0.5,angle=-90]{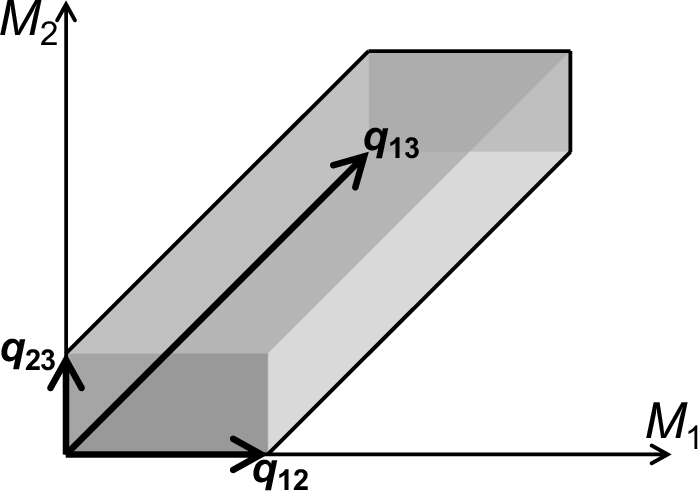}\quad\rotatebox[origin=c]{45}{\Huge =}\\[6pt]
(${\cal Z}$)
\caption{Three descriptions for the fundamental domain of duality cascades for the $\widehat A_2$ quiver.
The ${\cal H}$ description utilizes the inequalities in \eqref{a2ineq} obtained from various expressions of the brane configuration in table \ref{a2h}, while the ${\cal V}$ description is the convex hull of the vertices of the brane configuration without relative ranks in table \ref{a2v}.
The ${\cal Z}$ description originates from the S-rule which states that the number of D3-branes between each pair of 5-branes is restricted by the level.
Here we do not distinguish the D3-branes stretching between different pairs as long as the total relative ranks are equal.
This implies that we project the parallelepiped into lower dimensions, which is nothing but one of the definitions of zonotopes.}
\label{afig}
\end{figure}

The second one (the ${\cal Z}$ description) utilizes the S-rule.
It was proposed \cite{HW} that the number of D3-branes stretching between two 5-branes with charges $(p_1,q_1)$ and $(p_2,q_2)$ is bounded by the absolute value of the determinant of the two 5-brane charges.
Namely, for the current example with three 5-branes \eqref{three5}, the relative ranks $(M_1,M_2)$ in the brane configuration $\langle\stackrel{q_1}{\bullet}M_1\stackrel{q_2}{\bullet}M_2\stackrel{q_3}{\bullet}\rangle$ can be expressed as a linear combination,
\begin{align}
\langle\stackrel{q_1}{\bullet}M_1\stackrel{q_2}{\bullet}M_2\stackrel{q_3}{\bullet}\rangle
=c_{12}\langle\stackrel{q_1}{\bullet}|q_{12}|\stackrel{q_2}{\bullet}0\stackrel{q_3}{\bullet}\rangle+c_{13}\langle\stackrel{q_1}{\bullet}|q_{13}|\stackrel{q_2}{\bullet}|q_{13}|\stackrel{q_3}{\bullet}\rangle+c_{23}\langle\stackrel{q_1}{\bullet}0\stackrel{q_2}{\bullet}|q_{23}|\stackrel{q_3}{\bullet}\rangle,
\label{Sbasis}
\end{align}
or more explicitly
\begin{align}
M_1=c_{12}|q_{12}|+c_{13}|q_{13}|,\quad
M_2=c_{13}|q_{13}|+c_{23}|q_{23}|,
\label{M123}
\end{align}
with coefficients $0\le c_{ij}\le 1$.

Note that the ${\cal Z}$ description implies that, for the given relative ranks, we do not distinguish which pair of 5-branes each D3-brane ends on as long as the sum of the numbers of D3-branes in each interval reproduces the given total relative ranks.
This suggests that the fundamental domain in the two-dimensional ${\bm M}=(M_1,M_2)$ plane is a projection from a three-dimensional parallelepiped
\begin{align}
[0,|q_{12}|]\times[0,|q_{13}|]\times[0,|q_{23}|],
\end{align}
whose edges are projected to the vectors in two dimensions (see figure \ref{afig}),
\begin{align}
{\bm q}_{12}=|q_{12}|(1,0),\quad
{\bm q}_{13}=|q_{13}|(1,1),\quad
{\bm q}_{23}=|q_{23}|(0,1).
\label{q123}
\end{align}
This is nothing but the definition of the zonotope generated by the vectors \eqref{q123},
\begin{align}
{\bm M}=c_{12}{\bm q}_{12}+c_{13}{\bm q}_{13}+c_{23}{\bm q}_{23},\quad 0\le c_{ij}\le 1.
\label{Mcq}
\end{align}

It is important to note that the S-rule does not depend on orders of 5-branes.
Namely, given any set of relative ranks $\cdots K\bullet L\circ M\cdots$, if we can assign the numbers of D3-branes between all pairs of 5-branes (including those connecting non-neighboring 5-branes by trespassing intermediate ones) so that it satisfies the S-rule, we can assign as well the numbers of D3-branes satisfying the S-rule for the set of relative ranks $\cdots K\circ K-L+M+|k|\bullet M\cdots$ obtained by the HW transition.

The independence of the S-rule from orders of 5-branes can be seen explicitly.
In the order of $\langle\stackrel{q_2}{\bullet}\stackrel{q_1}{\bullet}\stackrel{q_3}{\bullet}\rangle$ obtained by exchanging 5-branes $\stackrel{q_1}{\bullet}\leftrightarrow\stackrel{q_2}{\bullet}$, the HW transition \eqref{hw} claims that the brane configuration \eqref{Sbasis} becomes
\begin{align}
&(1-c_{12})\langle\stackrel{q_2}{\bullet}|q_{12}|\stackrel{q_1}{\bullet}0\stackrel{q_3}{\bullet}\rangle+c_{13}\langle\stackrel{q_2}{\bullet}0\stackrel{q_1}{\bullet}|q_{13}|\stackrel{q_3}{\bullet}\rangle+c_{23}\langle\stackrel{q_2}{\bullet}|q_{23}|\stackrel{q_1}{\bullet}|q_{23}|\stackrel{q_3}{\bullet}\rangle\nonumber\\
&=
(1-c_{12})\langle\stackrel{q'_1}{\bullet}|q'_{12}|\stackrel{q'_2}{\bullet}0\stackrel{q'_3}{\bullet}\rangle+c_{13}\langle\stackrel{q'_1}{\bullet}0\stackrel{q'_2}{\bullet}|q'_{23}|\stackrel{q'_3}{\bullet}\rangle+c_{23}\langle\stackrel{q'_1}{\bullet}|q'_{13}|\stackrel{q'_2}{\bullet}|q'_{13}|\stackrel{q'_3}{\bullet}\rangle,
\end{align}
where we further reinterpret the charges by $(q'_1,q'_2,q'_3)=(q_2,q_1,q_3)$ and introduce $q'_{ij}=q'_i-q'_j$.
With the identification of coefficients
\begin{align}
c'_{12}=1-c_{12},\quad
c'_{23}=c_{13},\quad
c'_{13}=c_{23},
\end{align}
still satisfying $0\le c'_{ij}\le 1$, the new configuration is given as
\begin{align}
\langle\stackrel{q'_1}{\bullet}M'_1\stackrel{q'_2}{\bullet}M'_2\stackrel{q'_3}{\bullet}\rangle
=c'_{12}\langle\stackrel{q'_1}{\bullet}|q'_{12}|\stackrel{q'_2}{\bullet}0\stackrel{q'_3}{\bullet}\rangle+c'_{23}\langle\stackrel{q'_1}{\bullet}0\stackrel{q'_2}{\bullet}|q'_{23}|\stackrel{q'_3}{\bullet}\rangle+c'_{13}\langle\stackrel{q'_1}{\bullet}|q'_{13}|\stackrel{q'_2}{\bullet}|q'_{13}|\stackrel{q'_3}{\bullet}\rangle,
\label{sruleorder}
\end{align}
formally identical to \eqref{Sbasis}.
Similarly, this works for the order of $\langle\stackrel{q_1}{\bullet}\stackrel{q_3}{\bullet}\stackrel{q_2}{\bullet}\rangle$ obtained by exchanging 5-branes $\stackrel{q_2}{\bullet}\leftrightarrow\stackrel{q_3}{\bullet}$.
Since all the other orders are generated by these two exchanges, we claim that the S-rule does not depend on orders of 5-branes.

For later convenience, let us repeat the argument by rewriting the transformations algebraically.
We first note that the above computation in \eqref{sruleorder} consists of two steps, the application of the HW transition \eqref{hw} and the reinterpretation of charges.
To apply the transformation directly to the vectors \eqref{q123}, we separate from the original HW transition \eqref{aweyl} the ``homogeneous'' part
\begin{align}
r_i^\text{H}: q_i\leftrightarrow q_{i+1}, N_i\to N_{i-1}-N_i+N_{i+1},
\label{homo}
\end{align}
(where $N_i\to N_{i-1}-N_i+N_{i+1}$ is responsible for the application of the HW transition while $q_i\leftrightarrow q_{i+1}$ is for the reinterpretation of charges) and obtain
\begin{align}
r_1^\text{H}({\bm q}_{12},{\bm q}_{13},{\bm q}_{23})=(-{\bm q}_{12},{\bm q}_{23},{\bm q}_{13}),\quad
r_2^\text{H}({\bm q}_{12},{\bm q}_{13},{\bm q}_{23})=({\bm q}_{13},{\bm q}_{12},-{\bm q}_{23}).
\end{align}
Then, the independence of the S-rule from orders of 5-branes can be rewritten as the invariance of the relative ranks \eqref{Mcq} under the Weyl reflections.
For the Weyl reflections $r_i$ \eqref{aweyl} (including both the homogeneous and inhomogeneous parts), we find
\begin{align}
r_1({\bm M})&=r^\text{H}_1(c_{12}{\bm q}_{12}+c_{13}{\bm q}_{13}+c_{23}{\bm q}_{23})+|q_{12}|(1,0)
=(1-c_{12}){\bm q}_{12}+c_{13}{\bm q}_{23}+c_{23}{\bm q}_{13},\nonumber\\
r_2({\bm M})&=r^\text{H}_2(c_{12}{\bm q}_{12}+c_{13}{\bm q}_{13}+c_{23}{\bm q}_{23})+|q_{23}|(0,1)
=c_{12}{\bm q}_{13}+c_{13}{\bm q}_{12}+(1-c_{23}){\bm q}_{23}.
\end{align}
In both cases, after the maps, the new coefficients $c'_{ij}$ still satisfy the same condition, $0\le c'_{ij}\le 1$.
Therefore, two expressions are identical after change of variables and the S-rule is invariant under the Weyl reflections.

\begin{table}[!t]
\centering
\begin{tabular}{l|l}
brane configurations & relative ranks\\\hline
$\langle\stackrel{q_1}{\bullet}\stackrel{q_2}{\bullet}\stackrel{q_3}{\bullet}\rangle$&$(0,0)$\\
$\langle\stackrel{q_2}{\bullet}\stackrel{q_1}{\bullet}\stackrel{q_3}{\bullet}\rangle=\langle\stackrel{q_1}{\bullet}|q_{12}|\stackrel{q_2}{\bullet}\stackrel{q_3}{\bullet}\rangle$&$(|q_{12}|,0)$\\
$\langle\stackrel{q_2}{\bullet}\stackrel{q_3}{\bullet}\stackrel{q_1}{\bullet}\rangle=\langle\stackrel{q_1}{\bullet}|q_{12}|+|q_{13}|\stackrel{q_2}{\bullet}|q_{13}|\stackrel{q_3}{\bullet}\rangle$&$(|q_{12}|+|q_{13}|,|q_{13}|)$\\
$\langle\stackrel{q_3}{\bullet}\stackrel{q_2}{\bullet}\stackrel{q_1}{\bullet}\rangle=\langle\stackrel{q_1}{\bullet}|q_{12}|+|q_{13}|\stackrel{q_2}{\bullet}|q_{13}|+|q_{23}|\stackrel{q_3}{\bullet}\rangle$&$(|q_{12}|+|q_{13}|,|q_{13}|+|q_{23}|)$\\
$\langle\stackrel{q_3}{\bullet}\stackrel{q_1}{\bullet}\stackrel{q_2}{\bullet}\rangle=\langle\stackrel{q_1}{\bullet}|q_{13}|\stackrel{q_2}{\bullet}|q_{13}|+|q_{23}|\stackrel{q_3}{\bullet}\rangle$&$(|q_{13}|,|q_{13}|+|q_{23}|)$\\
$\langle\stackrel{q_1}{\bullet}\stackrel{q_3}{\bullet}\stackrel{q_2}{\bullet}\rangle=\langle\stackrel{q_1}{\bullet}\stackrel{q_2}{\bullet}|q_{23}|\stackrel{q_3}{\bullet}\rangle$&$(0,|q_{23}|)$
\end{tabular}
\caption{The relative ranks in the standard order, for brane configurations without relative ranks but in various orders of 5-branes.}
\label{a2v}
\end{table}

The third one (the ${\cal V}$ description) utilizes the brane configurations without relative ranks.
Namely, we assume that vertices of the fundamental domain are given by those configurations without relative ranks in various orders of 5-branes and define the fundamental domain by their convex hull.
As in table \ref{a2v}, for the configurations without relative ranks in various orders, we bring 5-branes into the standard order of $\langle\stackrel{q_1}{\bullet}\stackrel{q_2}{\bullet}\stackrel{q_3}{\bullet}\rangle$ and read off the relative ranks $(M_1,M_2)$ of the vertices.

As proved in \cite{FMS} and sketched below, all of these three descriptions are equivalent.
First we note that all of the three descriptions are apparently convex.
Since the S-rule does not depend on orders of 5-branes as we mentioned earlier, the ${\cal Z}$ description connects the ${\cal H}$ description on one hand and the ${\cal V}$ description on the other hand, ${\cal V}\subset{\cal Z}\subset{\cal H}$.
Namely, the vertices in the ${\cal V}$ description with vanishing relative ranks are elements in the ${\cal Z}$ description and the ${\cal Z}$ description guarantees all the ranks to satisfy the inequalities in the ${\cal H}$ description.

The reverse subset relation is more non-trivial (see \cite{FMS} for details).
For simplicity we only provide an intuitive picture here.
Note that each order of 5-branes induces a set of inequalities which forms a cone locally with the tip being a vertex of the polytope of the fundamental domain and the surrounding hypersurfaces being the codimension-one facets intersecting at the vertex.
In applying one of the HW transitions or the Weyl reflections, since only one of the inequalities changes, we are moving along an edge.
By requiring the vanishing of the new rank appearing from the Weyl reflection, we encounter the other vertex on the edge.
Thus, we can relate directly the ${\cal H}$ description to the ${\cal V}$ description.

\begin{table}[!t]
\centering
\begin{tabular}{c|c|c}
inequalities&shifts $\Delta N$&translations $\Delta{\bm M}$\\\hline
$M_1\ge 0$&$M_1$&$\Delta_1{\bm M}$\\
$M_2\ge 0$&$M_2$&$\Delta_2{\bm M}$\\
$-M_1+M_2+|q_{12}|\ge 0$&$-M_1+M_2+|q_{12}|$&$-\Delta_1{\bm M}+\Delta_2{\bm M}$\\
$-M_1+|q_{12}|+|q_{13}|\ge 0$&$-M_1+|q_{12}|+|q_{13}|$&$-\Delta_1{\bm M}$\\
$-M_2+|q_{13}|+|q_{23}|\ge 0$&$-M_2+|q_{13}|+|q_{23}|$&$-\Delta_2{\bm M}$\\
$M_1-M_2+|q_{23}|\ge 0$&$M_1-M_2+|q_{23}|$&$\Delta_1{\bm M}-\Delta_2{\bm M}$
\end{tabular}
\caption{The shift of $N$, $\Delta N=N'-N$, and the translation of ${\bm M}$, $\Delta{\bm M}={\bm M}'-{\bm M}$, which are induced as one step of duality cascades when the inequality is violated.}
\label{a2t}
\end{table}

Out of the three descriptions, the ${\cal Z}$ description is the most useful.
It was known in \cite{Sh,M} that the space-tiling condition for a zonotope is that the rank of the facet centers is equal to the space dimension.
Since the facet centers correspond to the discrete translations in tiling the space, this condition is equivalent to the condition that all the discrete translations are compatible in the space.
In \cite{FMS} it was shown that, for an arbitrary number of 5-branes with arbitrary charges, the space-tiling condition is satisfied.
This implies that the fundamental domain is a parallelotope.
Hence, the questions on the finiteness and the uniqueness are answered positively.

\begin{figure}[t!]
\centering\includegraphics[scale=0.5,angle=-90]{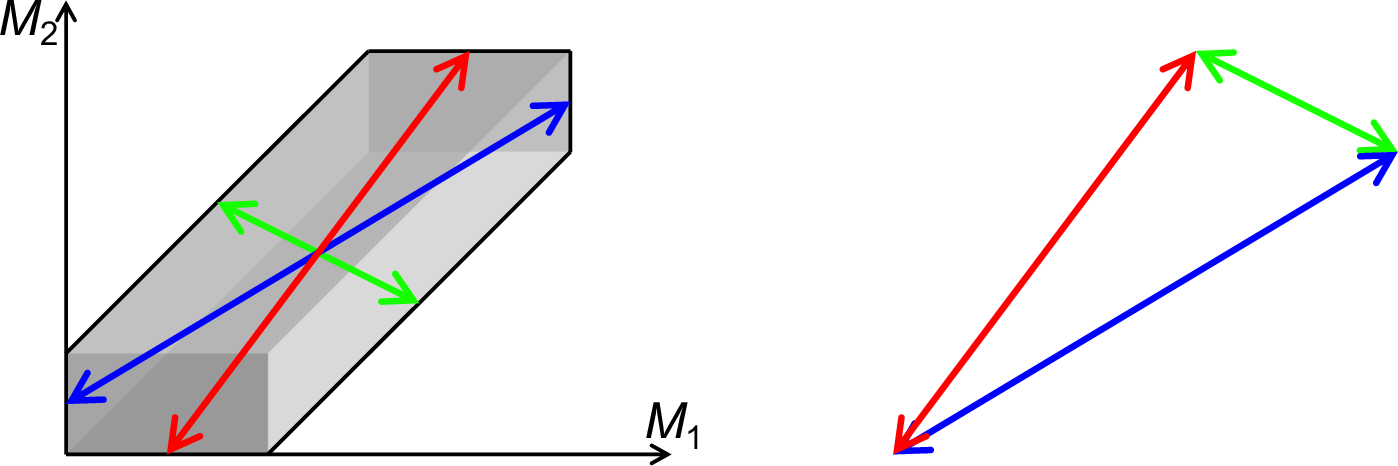}
\caption{Compatibility of the discrete translations of duality cascades.
Three discrete translations for the brane configuration \eqref{three5} are compatible in two dimensions.}
\label{tiling}
\end{figure}

Let us see explicitly the discrete translations of duality cascades and their compatibility in our previous example with three 5-branes \eqref{three5}.
Suppose that $N+M_1$ is smaller than the reference rank $N$, we change references as $\langle N+M_1\stackrel{q_2}{\bullet} N+M_2\stackrel{q_3}{\bullet}N\stackrel{q_1}{\bullet}\rangle$.
By reordering the 5-branes into the standard one $\langle \stackrel{q_1}{\bullet}\stackrel{q_2}{\bullet}\stackrel{q_3}{\bullet}\rangle$ using the HW transitions, we find $\langle N'\stackrel{q_1}{\bullet}N'+M'_1\stackrel{q_2}{\bullet} N'+M'_2\stackrel{q_3}{\bullet}\rangle$ with
\begin{align}
N'=N+M_1,\quad M'_1=M_1+|q_{12}|+|q_{13}|,\quad M'_2=M_2+|q_{13}|.
\end{align}
Thus, the transformation for the relative ranks ${\bm M}=(M_1,M_2)$ is realized as a discrete translation in the parameter space of relative ranks $(M_1,M_2)$.
Similarly, for other inequalities, transformations under duality cascades are also realized as discrete translations, as summarized in table \ref{a2t} with
\begin{align}
\Delta_1{\bm M}={\bm q}_{12}+{\bm q}_{13},\quad
\Delta_2{\bm M}={\bm q}_{13}+{\bm q}_{23}.
\label{delta12m}
\end{align}
It is non-trivial that three pairs of discrete translations $\Delta{\bm M}$ in two dimensions can be expressed in terms of the two in \eqref{delta12m} with integral coefficients.
This is why all the discrete translations are compatible in two dimensions and hence the fundamental domain forms a parallelotope.
See figure \ref{tiling} for the compatibility in the current setup.

\subsection{More group-theoretical analysis}

\begin{figure}[t!]
\centering\includegraphics[scale=0.6,angle=-90]{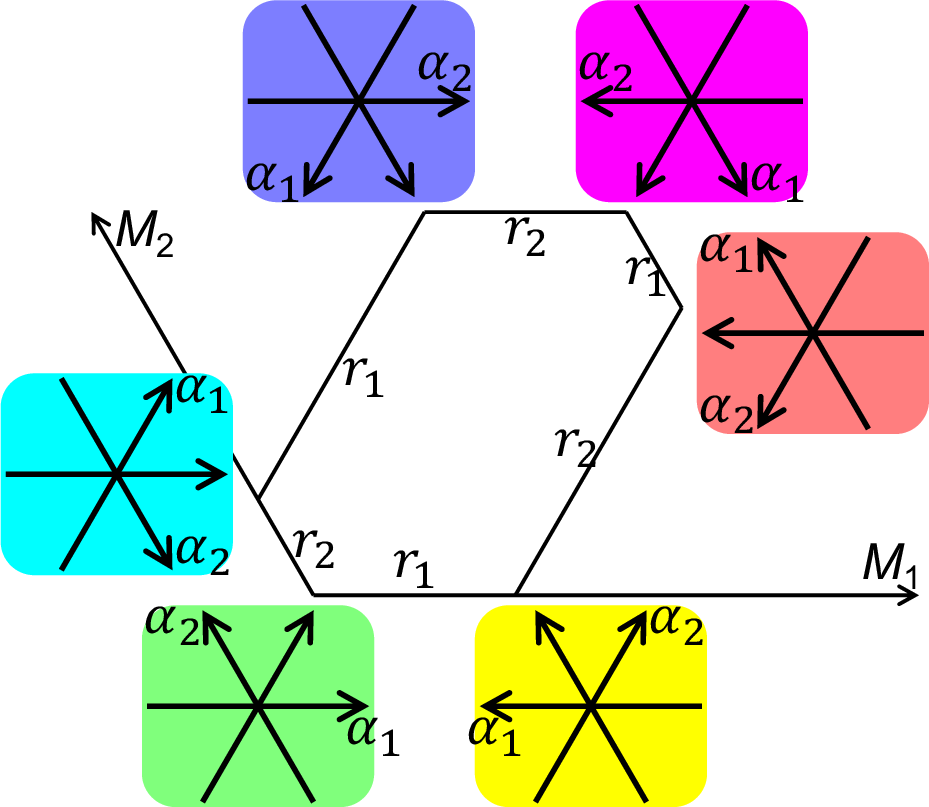}
\caption{The relation between vectors of the zonotope and positive roots in the root system.}
\label{vectorroot}
\end{figure}

We have interpreted the duality transformations of the HW transitions \eqref{hw} as the Weyl reflections \eqref{aweyl} and explained duality cascades by relating to the Weyl group.
Since the Weyl group is originally defined as symmetries for the root system, it is interesting to clarify the relation.
For this purpose, we depict the fundamental domain with the axes tilted (see figure \ref{vectorroot}) and match the vectors of the zonotope with the positive roots.
Namely, we first concentrate on the cone around the vertex at the origin $(0,0)$ and identify vectors for the zonotope ${\bm q}_{12}$, ${\bm q}_{13}$, ${\bm q}_{23}$ as positive roots for the root system $\alpha_1$, $\alpha_1+\alpha_2$, $\alpha_2$ respectively.
Then, we find a similarity in the homogeneous part of the HW transition $r_i^\text{H}$ \eqref{homo} and the usual Weyl reflection $r_i$ for the positive roots.

Indeed, by applying the homogeneous part of the HW transition $r^\text{H}_1$ for the vectors and the Weyl reflection $r_1$ for the positive roots, we find isomorphic relations
\begin{align}
&({\bm q}'_{12},{\bm q}'_{13},{\bm q}'_{23})
=r^\text{H}_1({\bm q}_{12},{\bm q}_{13},{\bm q}_{23})
=(-{\bm q}_{12},{\bm q}_{23},{\bm q}_{13}),\nonumber\\
&(\alpha'_1,\alpha'_1+\alpha'_2,\alpha'_2)
=r_1(\alpha_1,\alpha_1+\alpha_2,\alpha_2)
=(-\alpha_1,\alpha_2,\alpha_1+\alpha_2),
\end{align}
respectively.
Subsequently, counterclockwise, the vectors and the positive roots are transformed respectively by $r^{\prime\text{H}}_2$ and $r'_2$ (defined similarly for vectors with primes) as
\begin{align}
&({\bm q}''_{12},{\bm q}''_{13},{\bm q}''_{23})
=r^{\prime\text{H}}_2({\bm q}'_{12},{\bm q}'_{13},{\bm q}'_{23})
=({\bm q}'_{13},{\bm q}'_{12},-{\bm q}'_{23})
=({\bm q}_{23},-{\bm q}_{12},-{\bm q}_{13}),\nonumber\\
&(\alpha''_1,\alpha''_1+\alpha''_2,\alpha''_2)
=r'_2(\alpha'_1,\alpha'_1+\alpha'_2,\alpha'_2)
=(\alpha'_1+\alpha'_2,\alpha'_1,-\alpha'_2)
=(\alpha_2,-\alpha_1,-\alpha_1-\alpha_2).
\end{align}
We can also continue by $r^{\prime\prime\text{H}}_1$ and $r''_1$ to find the transformation
\begin{align}
&({\bm q}'''_{12},{\bm q}'''_{13},{\bm q}'''_{23})
=r^{\prime\prime\text{H}}_1({\bm q}''_{12},{\bm q}''_{13},{\bm q}''_{23})
=(-{\bm q}''_{12},{\bm q}''_{23},{\bm q}''_{13})
=(-{\bm q}_{23},-{\bm q}_{13},-{\bm q}_{12}),\nonumber\\
&(\alpha'''_1,\alpha'''_1+\alpha'''_2,\alpha'''_2)
=r''_1(\alpha''_1,\alpha''_1+\alpha''_2,\alpha''_2)
=(-\alpha''_1,\alpha''_2,\alpha''_1+\alpha''_2)
=(-\alpha_2,-\alpha_1-\alpha_2,-\alpha_1).
\end{align}
Namely, the vectors of the zonotope which specify the directions the vertex extends to in the cone serve the role of the positive roots in the root system.
On the other hand, the inhomogeneous part of the HW transition \eqref{aweyl} induces the transformation from the origin $(0,0)$ subsequently into $(|q_{12}|,0)$, $(|q_{12}|+|q_{13}|,|q_{13}|)$ and $(|q_{12}|+|q_{13}|,|q_{13}|+|q_{23}|)$.
In figure \ref{vectorroot}, we see explicitly that the tips and the vectors of the zonotope change counterclockwise.
To summarize, the homogeneous part of the HW transition $r_i^\text{H}$ \eqref{homo} is isomorphic to the Weyl reflection on the root vectors, while the inhomogeneous part induces the change of the vertices.

So far we have seen that vectors of the zonotope correspond to positive roots of the root system, which implies that their numbers are identical.
Various other geometric quantities of the polytope of the fundamental domain can also be evaluated.
In the ${\cal H}$ description, the facets are identified with inequalities, which are characterized by separating 5-branes into two disjoint non-empty subsets from the charge conservation $Q_\text{RR}$.
This implies that the polytope has $2^{n+1}-2$ facets.
In the ${\cal V}$ description, the vertices are identified with brane configurations in various orders without relative ranks.
This implies that the polytope has $(n+1)!$ vertices.

Besides, the evaluation of the numbers of facets can be refined from the group-theoretical viewpoint.
In the example of $n=2$ with three 5-branes in figure \ref{afig}, we have $3!=6$ vertices and $2^3-2=6$ facets.
Since facets are characterized by the separation of 5-branes, if we denote 5-branes by $\bullet$ and the separation by $\times$, the facets may be expressed by $\langle{\bullet}{\times}{\bullet}{\bullet}\rangle$ and $\langle{\bullet}{\bullet}{\times}{\bullet}\rangle$.
The former contains $3!/(1!2!)=3$ facets, $M_1\ge 0$ from $\langle\stackrel{1}{\bullet}{\times}\stackrel{2}{\bullet}\stackrel{3}{\bullet}\rangle$, $|q_{12}|-M_1+M_2\ge 0$ from
$\langle\stackrel{2}{\bullet}{\times}\stackrel{1}{\bullet}\stackrel{3}{\bullet}\rangle$ and $|q_{13}|+|q_{23}|-M_2\ge 0$ from $\langle\stackrel{3}{\bullet}{\times}\stackrel{1}{\bullet}\stackrel{2}{\bullet}\rangle$, while the latter contains $3!/(2!1!)=3$ facets, $M_2\ge 0$ from $\langle\stackrel{1}{\bullet}\stackrel{2}{\bullet}{\times}\stackrel{3}{\bullet}\rangle$, $|q_{23}|+M_1-M_2\ge 0$ from $\langle\stackrel{1}{\bullet}\stackrel{3}{\bullet}{\times}\stackrel{2}{\bullet}\rangle$ and $|q_{12}|+|q_{13}|-M_1\ge 0$ from $\langle\stackrel{2}{\bullet}\stackrel{3}{\bullet}{\times}\stackrel{1}{\bullet}\rangle$.
Each facet has two vertices.
This also works for the $n=3$ case with four 5-branes studied carefully in \cite{FMS}.
This time we have $4!/(1!3!)=4$ facets of type $\langle{\bullet}{\times}{\bullet}{\bullet}{\bullet}\rangle$ having $1!3!=6$ vertices, $4!/(2!2!)=6$ facets of type $\langle{\bullet}{\bullet}{\times}{\bullet}{\bullet}\rangle$ having $2!2!=4$ vertices and $4!/(3!1!)=4$ facets of type $\langle{\bullet}{\bullet}{\bullet}{\times}{\bullet}\rangle$ having $3!1!=6$ vertices.
Indeed, the polytope of the fundamental domain for the $n=3$ case is the truncated octahedron having $4+4=8$ hexagons and $6$ parallelograms for its facets (see figures in \cite{FMS}).

As is clear by now, brane configurations are not mandatory in these arguments.
The Dynkin diagram serves a similar but more crucial role.
Considering facets of codimension one corresponds to removing one node from the Dynkin diagram (as what we usually do in studying maximal subalgebras).
For example, for the above case of $A_2$, \textcircled{\scriptsize 1}--\textcircled{\scriptsize 2}, with three 5-branes \eqref{three5}, by removing one node, we have \textcircled{\scriptsize 1} or \textcircled{\scriptsize 2}.
Here and hereafter, we abbreviate the gauge group factor $\text{U}(N_i)_{k_i}$ in $\text{U}(N_0)_{k_0}\times\text{U}(N_1)_{k_1}\times\cdots\times\text{U}(N_n)_{k_n}$ simply as a node \textcircled{\scriptsize $i$}.
For the case of four nodes studied in \cite{FMS} (corresponding to $A_3$, \textcircled{\scriptsize 1}--\textcircled{\scriptsize 2}--\textcircled{\scriptsize 3}), by removing one node, we have \textcircled{\scriptsize 2}--\textcircled{\scriptsize 3}, \textcircled{\scriptsize 1} \textcircled{\scriptsize 3} or \textcircled{\scriptsize 1}--\textcircled{\scriptsize 2}.
Generally, by labelling \textcircled{\scriptsize 1}--\textcircled{\scriptsize 2}--\textcircled{\scriptsize 3}--$\cdots$--\textcircled{\scriptsize $n$} for the $A_n$ Dynkin diagram, the facets can be studied by removing any of the nodes \textcircled{\scriptsize $j$}, which reduces the Dynkin diagram into that of $A_{j-1}\times A_{n-j}$ containing $(n+1)!/(j!(n-j+1)!)$ facets.
Summing them up, we find
\begin{align}
\sum_{j=1}^n\frac{(n+1)!}{j!(n-j+1)!}=2^{n+1}-2,
\label{facetsumAn}
\end{align}
matching with the total number of facets evaluated in the above paragraphs.

To summarize, in this section, we have rewritten duality cascades for the circular quivers into the group-theoretical language and understand various properties from the Weyl group.
We have found that, as long as we have the affine Dynkin diagram, the original brane configuration seems not very important.
We have found that the working hypothesis of duality cascades can be stated purely in terms of the group-theoretical language.
Also, the three descriptions of the fundamental domain and its geometric properties are all understood with the group theory.
After the rewriting, it is interesting to generalize duality cascades for other quivers of affine Dynkin diagrams, such as $\widehat D$ and $\widehat E$.
This is the main subject we shall turn to in the subsequent sections.

It was also noted \cite{FMS} that the fundamental domain is equivalent to the permutohedron, a convex hull of all permutations of the vertex coordinates, since the fundamental domain has the ${\cal V}$ description with the vertices being brane configurations without relative ranks in various orders.
It is interesting to see what polytopes we encounter in generalizing the super Chern-Simons theories of the $\widehat A$ quivers into other affine quivers such as the $\widehat D$ and $\widehat E$ quivers.

\section{$\widehat D$ quivers}\label{D}

In the previous section, we have recapitulated duality cascades for the super Chern-Simons theories of circular quivers or the $\widehat A$ quivers in the group-theoretical language.
After rewriting into groups, the generalization to other simply-laced quivers should not be difficult.
In this section, we first study the $\widehat D_4$ quiver in details, then shortly turn to $\widehat D_5$ and more general $\widehat D_n$ quivers.

\subsection{$\widehat D_4$ quiver}

\begin{figure}[t!]
\centering\includegraphics[scale=0.5,angle=-90]{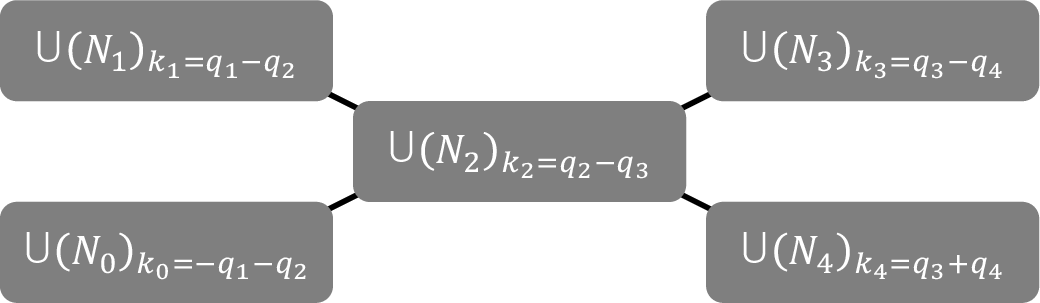}
\caption{$\widehat D_4$ quiver for the super Chern-Simons theory.
As in figure \ref{An}, lines connecting nodes $\text{U}(N_i)_{k_i}$ and $\text{U}(N_j)_{k_j}$ denote pairs of bifundamental matters $(N_i,\overline N_j)$ and  $(\overline N_i,N_j)$.}
\label{d4}
\end{figure}

Let us start with the simplest generalization, the quiver gauge theory of type $\widehat D_4$.
Namely, we consider the ${\cal N}=3$ super Chern-Simons theory of gauge group $\text{U}(N_0)_{k_0}\times\text{U}(N_1)_{k_1}\times\text{U}(N_2)_{k_2}\times\text{U}(N_3)_{k_3}\times\text{U}(N_4)_{k_4}$ with bifundamental matters specified by lines in the quiver diagram given in figure \ref{d4}.
While levels are subject to the zero sum condition $\sum_{i=0}^nk_i=0$ previously for the $\widehat A$ quivers, for the $\widehat D_4$ quiver levels are subject to $k_0+k_1+2k_2+k_3+k_4=0$ with the non-trivial mark $2$.
The physical partition functions were recently studied extensively in \cite{KN}.

\subsubsection{Duality cascades}

The HW transitions are originally defined for brane configurations and the HW transitions for other general affine quivers may be ambiguous.
Despite the ambiguities, after reinterpreting HW transitions \eqref{hw} as Weyl reflections \eqref{aweyl} for the $\widehat A$ quivers, we can easily guess the duality transformations of levels, since the transformations of simple roots under the Weyl reflections are well-known.
Also, since ranks of the nodes adjacent to the node of the Weyl reflection serve uniformly as bifundamental matters in gauge theories, the homogeneous transformations of ranks should be the difference between the sum of the adjacent ranks and the original rank, analogously as the $\widehat A$ quivers.
Namely, inspired by the Weyl reflections, we propose the duality transformations
\begin{align}
r_1&:k_1\to-k_1,\;k_2\to k_2+k_1,\;N_1\to N_2-N_1+|k_1|,\nonumber\\
r_2&:k_2\to-k_2,\;k_0\to k_0+k_2,\;k_1\to k_1+k_2,\;k_3\to k_3+k_2,\;k_4\to k_4+k_2,\nonumber\\
&\hspace{40mm}N_2\to N_0+N_1+N_3+N_4-N_2+|k_2|,\nonumber\\
r_3&:k_3\to-k_3,\;k_2\to k_2+k_3,\;N_3\to N_2-N_3+|k_3|,\nonumber\\
r_4&:k_4\to-k_4,\;k_2\to k_2+k_4,\;N_4\to N_2-N_4+|k_4|.
\label{d4weyl}
\end{align}
Here the affine Weyl reflection $r_0$ is given by
\begin{align}
r_0&:k_0\to-k_0,\;k_2\to k_2+k_0,\;N_0\to N_2-N_0+|k_0|,
\end{align}
which we do not use after we fix the reference in duality cascades.
From the analogy to the original Weyl reflections \eqref{weylrefl}, it is obvious that these duality transformations of levels $k_i$ satisfy the same algebra, $r_i^2=1$, $(r_ir_j)^3=1$ for connected pairs of nodes $(i,j)$ and $(r_ir_j)^2=1$ for disconnected pairs.
Besides, it is interesting to find that the algebra is not ruined by the transformations of ranks $N_i$.
It is of course an important issue to verify these duality transformations as gauge theories, or at least for their partition functions localized to matrix models.
Here we assume the duality transformations \eqref{d4weyl} and postpone the justifications from the viewpoint of gauge theories to future studies.
Hereafter we often parameterize the levels by
\begin{align}
k_0=-q_1-q_2,\quad
k_1=q_1-q_2,\quad
k_2=q_2-q_3,\quad
k_3=q_3-q_4,\quad
k_4=q_3+q_4.
\label{kq}
\end{align}
In terms of the variables $(q_1,q_2,q_3,q_4)$, the transformations of levels are given by
\begin{align}
r_1:q_1\leftrightarrow q_2,\quad
r_2:q_2\leftrightarrow q_3,\quad
r_3:q_3\leftrightarrow q_4,\quad
r_4:q_3\leftrightarrow -q_4.
\end{align}
Also, let us parameterize the ranks by
\begin{align}
N_0=N,\quad N_1=N+M_1,\quad N_2=2N+M_2,\quad N_3=N+M_3,\quad N_4=N+M_4,
\label{d4parameter}
\end{align}
so that the dependence on the ``overall'' rank $N$ does not change under the transformations $r_i$.
The parameterization with a coefficient $2$ in $N_2$ is related to the non-trivial mark.

It is known that, for the $\widehat D_4$ affine Dynkin diagram, we have two independent outer automorphisms ${\mathbb Z}_2\times{\mathbb Z}_2$,
\begin{align}
\sigma_1&:k_0\leftrightarrow k_1,\;k_3\leftrightarrow k_4,\;N_0\leftrightarrow N_1,\;N_3\leftrightarrow N_4,\nonumber\\
\sigma_2&:k_0\leftrightarrow k_4,\;k_1\leftrightarrow k_3,\;N_0\leftrightarrow N_4,\;N_1\leftrightarrow N_3,
\label{d4outer}
\end{align}
which can be used to map finite nodes into the affine one, when the finite nodes have lower ranks compared with the affine one.

With the Weyl reflections \eqref{d4weyl} and the outer automorphisms \eqref{d4outer}, first it is interesting to observe duality cascades for the super Chern-Simons theory of the $\widehat D_4$ quiver. 
For general quivers of the affine Dynkin diagrams (even other than the $\widehat A$ quivers), we adopt the same working hypothesis given below \eqref{aweyl}.
\begin{itemize}
\item[$1'.$] We apply finite Weyl reflections arbitrarily (without using the affine one).  
\item[$2'.$] We change affine nodes by applying the outer automorphisms of the affine Weyl group, when encountering lower ranks compared with that of the current affine node.
\end{itemize}

\begin{figure}[t!]
\centering\includegraphics[scale=0.5,angle=-90]{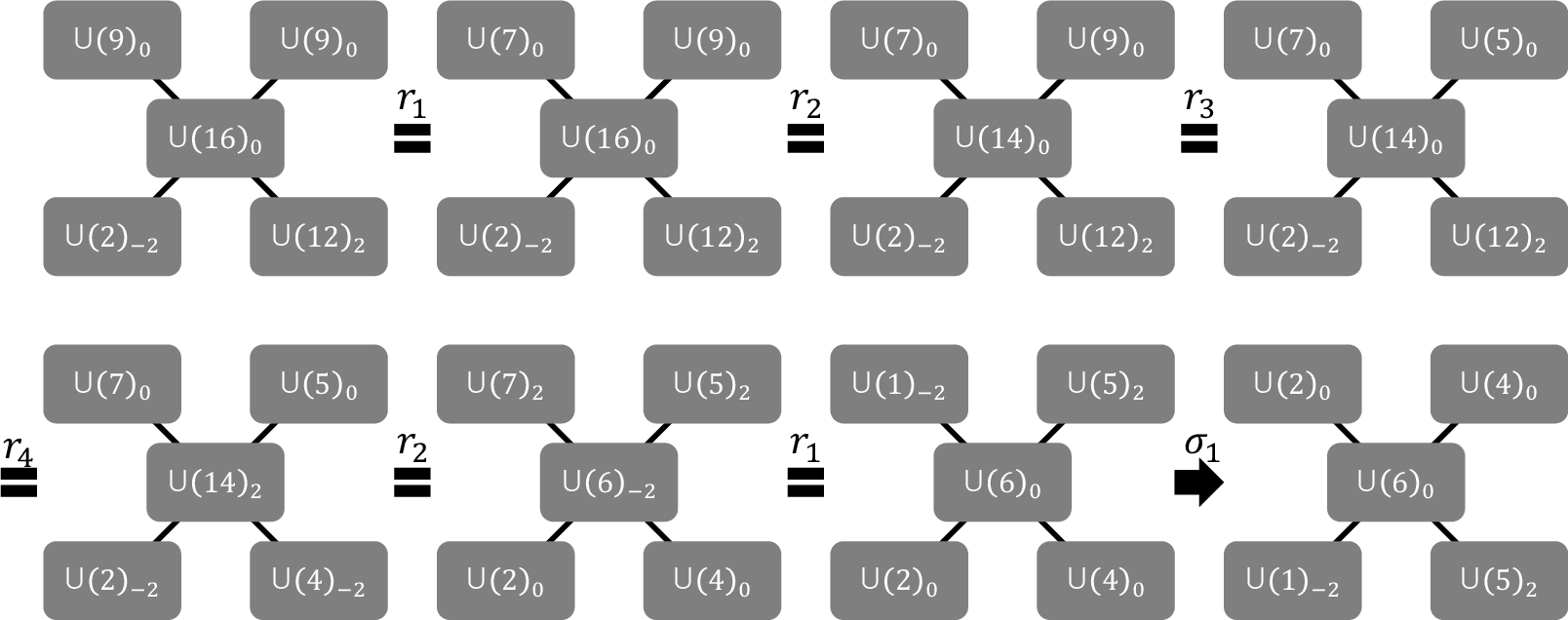}
\caption{An example of duality cascades for the super Chern-Simons theory of the $\widehat D_4$ quiver.
After applying a sequence of Weyl reflections $r_1r_2r_4r_3r_2r_1$, we encounter a rank lower than that of the affine node.
Then, we can use the outer automorphism $\sigma_1$ to change affine nodes.}
\label{d4dualitycascade}
\end{figure}

As an example, let us consider duality cascades for the $\widehat D_4$ quiver with gauge group $\text{U}(2)_{-2}\times\text{U}(9)_0\times\text{U}(16)_0\times\text{U}(9)_0\times\text{U}(12)_2$ and bifundamental matters indicated by the $\widehat D_4$ quiver by setting $N=2,(M_1,M_2,M_3,M_4)=(7,12,7,10),q_1=q_2=q_3=q_4=1$ in figure \ref{d4} (see table \ref{d4dualitycascade}).
Note that the rank of the affine root is the lowest.
By applying the Weyl reflections $r_1r_2r_4r_3r_2r_1$ subsequently, we find the gauge group finally reduces to $\text{U}(2)_0\times\text{U}(1)_{-2}\times\text{U}(6)_0\times\text{U}(5)_2\times\text{U}(4)_0$.
Now since the rank of the original affine node is not the lowest, let us apply the outer automorphism $\sigma_1$ to bring the node of the lowest rank to the affine node.
Finally we find $\text{U}(1)_{-2}\times\text{U}(2)_0\times\text{U}(6)_0\times\text{U}(4)_0\times\text{U}(5)_2$.
Note that, although some of the Weyl reflections are unnecessary for the current choice of $q_1=q_2=q_3=q_4$, we adopt them for later studies of general cases.

Here, after the subsequent Weyl reflections $r_1r_2r_4r_3r_2r_1$ accompanied by the outer automorphism $\sigma_1$, the original ranks decrease with the levels being identical.
Hence, it deserves the name of duality cascades.
Although sometimes we can continue duality cascades by more steps, the current gauge theory does not reduce any more.
Note that duality cascades occur only partially.
Since we need to return to the original levels with the outer automorphism ${\mathbb Z}_2\times{\mathbb Z}_2$ \eqref{d4outer}, duality cascades can only occur when the ranks in the nodes \textcircled{\scriptsize 1}, \textcircled{\scriptsize 3}, \textcircled{\scriptsize 4} are lower than that of the affine node \textcircled{\scriptsize 0}, while we cannot compare with the rank in the node \textcircled{\scriptsize 2}.
For this reason, we shall distinguish the nodes \textcircled{\scriptsize 1}, \textcircled{\scriptsize 3}, \textcircled{\scriptsize 4} as the potentially affine nodes from the non-affine node \textcircled{\scriptsize 2}.
Here, as for the $\widehat A$ quivers, we denote the gauge group factor $\text{U}(N_i)_{k_i}$ in figure \ref{d4} by \textcircled{\scriptsize $i$}.
Note that this process of duality cascades is reminiscent of the process of finding discrete translations discussed in \cite{KNY}.
In other words, we may want to claim that the process in \cite{KNY} has been given a physical interpretation from duality cascades.

As in the previous case for the $\widehat A$ quivers, the most crucial questions to study would be, adopting the working hypothesis of duality cascades given above, whether duality cascades always terminate in finite steps and whether the endpoint of duality cascades is uniquely determined depending only on a starting point regardless of processes of duality cascades.
As before, we would like to study these questions by defining the fundamental domain of duality cascades.
Previously for the $\widehat A$ quivers, we have three descriptions of them, all of which are proved to be equivalent, and the space-tiling condition was confirmed using the ${\cal Z}$ description.
Here we still expect that similar arguments hold, except for the group-theoretical difference between the $\widehat A$ quivers and the $\widehat D$ and $\widehat E$ quivers.
First, as explained above, not all nodes can be the affine node for the $\widehat D\widehat E$ quivers.
By applying the outer automorphism, we can map only some nodes into the affine node, which should be distinguished from the others.
Second, the $\widehat D\widehat E$ quivers have non-trivial marks.
We will see at the end of this subsection how the non-trivial marks disturb the space-tiling condition.

\subsubsection{Three descriptions}

To answer the questions on the finiteness and the uniqueness raised above, as in the $\widehat A$ quivers, we attempt to provide three descriptions of the fundamental domain of duality cascades with the parameterization \eqref{d4parameter}. 

The ${\cal H}$ description of the fundamental domain for the $\widehat A$ quivers was defined to be the combination of inequalities comparing the rank of the affine node with all the other ranks in the quiver diagram obtained by the HW transitions (or the finite Weyl reflections).
In the ${\cal H}$ description for the current $\widehat D_4$ quiver, by requiring that, under the applications of the finite Weyl reflections \eqref{d4weyl}, the rank of the affine node should stay the lowest,
\begin{align}
N_0\le N'_1,\quad 2N_0\le N'_2,\quad N_0\le N'_3,\quad N_0\le N'_4,
\end{align}
we find 48 inequalities.
Here $(N'_1,N'_2,N'_3,N'_4)$ are the ranks at the nodes \textcircled{\scriptsize 1},\textcircled{\scriptsize 2},\textcircled{\scriptsize 3},\textcircled{\scriptsize 4} respectively obtained by the Weyl reflections.
Namely, by applying the finite Weyl reflections \eqref{d4weyl} to the original parameterization \eqref{d4parameter}, we obtain various ranks for the finite nodes, which we require to stay larger than that for the affine one.
Here 24 originate from the comparison of the affine node \textcircled{\scriptsize 0} with the potentially affine nodes \textcircled{\scriptsize 1}, \textcircled{\scriptsize 3}, \textcircled{\scriptsize 4},
\begin{align}
&{\pm M_i}+\cdots\ge 0,\quad
\pm(M_2-M_i)+\cdots\ge 0,\quad
\pm(M_i+M_j-M_2)+\cdots\ge 0,\nonumber\\
&(M_i-M_j)+\cdots\ge 0,
\label{d4affineineq}
\end{align}
while the remaining 24 are from the other node \textcircled{\scriptsize 2}
\begin{align}
&{\pm M_2}+\cdots\ge 0,\quad\pm(M_2-2M_i)+\cdots\ge 0,\quad
\pm(M_i+M_j-M_k)+\cdots\ge 0,\nonumber\\
&{\pm(M_i+M_j+M_k-M_2)}+\cdots\ge 0,\quad\pm(M_2-M_i-M_j+M_k)+\cdots\ge 0,\nonumber\\
&{\pm(2M_2-M_i-M_j-M_k)}+\cdots\ge 0,
\label{d4nonaffineineq}
\end{align}
with $i,j,k=1,3,4$ being different.
Here in \eqref{d4affineineq} and \eqref{d4nonaffineineq} we have omitted the absolute values of the levels, which are recorded explicitly later in tables \ref{D4affine} and \ref{D4nonaffine}.
The number of facets can also be understood later from the group theory.

Note that the word, fundamental domain, may be misleading.
For the $\widehat A$ quivers, the parameter domain where duality cascades no more occur is called the fundamental domain of duality cascades because it serves as the set of all representative configurations under duality cascades.
For other quivers, as we mention in the example in figure \ref{d4dualitycascade}, the comparison of the rank of the affine node with those of the non-affine nodes does not induce duality cascades.
For this reason, we often avoid the misleading word and just call it the polytope associated to duality cascades.

The ${\cal V}$ description of the fundamental domain for the $\widehat A$ quivers was defined to be the convex hull of the configuration in various orders of 5-branes without relative ranks.
Similarly in the ${\cal V}$ description for the current $\widehat D_4$ quiver, we consider various configurations of levels generated from the finite Weyl reflections \eqref{d4weyl} but with relative ranks set to zero and then bring them back to the original configurations of levels using the finite Weyl reflections \eqref{d4weyl} to read off the relative ranks.
Totally the number of vertices is the order of the $D_4$ Weyl group $\#(W(D_4))=192$, since they are generated by the Weyl group.

\begin{figure}[t!]
\centering\includegraphics[scale=0.5,angle=-90]{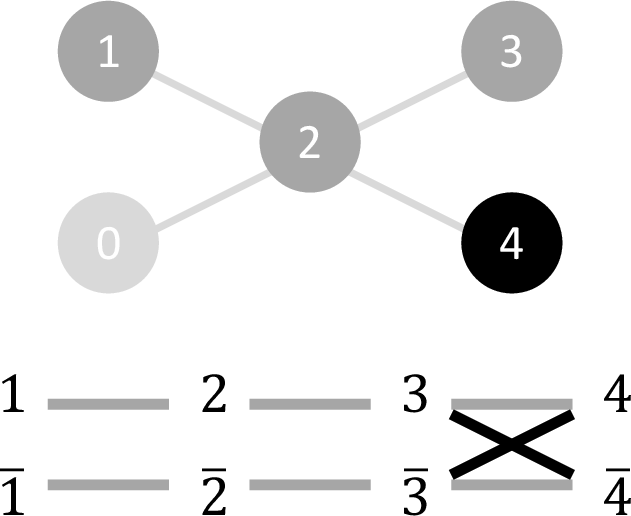}
\caption{S-rule for the $\widehat D_4$ quiver.
The original S-rule stating that the numbers of D3-branes connecting two 5-branes $i$ and $j$ are bounded by $|q^-_{ij}|$ via the Weyl reflections $r_1=(1\leftrightarrow 2,\bar 1\leftrightarrow\bar 2)$, $r_2=(2\leftrightarrow 3,\bar 2\leftrightarrow\bar 3)$, $r_3=(3\leftrightarrow 4,\bar 3\leftrightarrow\bar 4)$ depicted in light grey.
Besides it, we can connect the 5-brane $i$ and the mirror image $\bar j$ with the numbers of D3-branes bounded by $|q^+_{ij}|$ using the Weyl reflection $r_4=(3\leftrightarrow\bar 4,\bar 3\leftrightarrow 4)$ depicted in dark grey.
For example for $|q^+_{12}|$, to connect $1\leftrightarrow\bar 2$ we have to connect by $1\leftrightarrow 2\leftrightarrow 3\leftrightarrow 4\leftrightarrow\bar 3\leftrightarrow\bar 2$ with $r_2r_4r_3r_2r_1$ giving rise to $\#(r_1,r_2,r_3,r_4)=(1,2,1,1)$.}
\label{SforD}
\end{figure}

The ${\cal Z}$ description for the $\widehat A$ quivers was constructed from the S-rule assuming that the number of D3-branes between two 5-branes is suppressed by the absolute value of the determinant of the two 5-brane charges.
Since the S-rule is not fully established for other quivers, we cannot adopt the definition directly.
Instead, let us take the ${\cal V}$ description and construct the zonotope in the ${\cal Z}$ description out of it.
Namely, by setting one of the variables $q_{ij}^\mp=q_i\mp q_j$ to zero, all the vertices in the ${\cal V}$ description either remain or reduce by a vector.
For each variable, the corresponding vector is given by
\begin{align}
&{\bm q}^-_{12}=|q^-_{12}|(1,0,0,0),\quad
{\bm q}^-_{13}=|q^-_{13}|(1,1,0,0),\quad
{\bm q}^-_{14}=|q^-_{14}|(1,1,1,0),\nonumber\\
&{\bm q}^-_{23}=|q^-_{23}|(0,1,0,0),\quad
{\bm q}^-_{24}=|q^-_{24}|(0,1,1,0),\quad
{\bm q}^-_{34}=|q^-_{34}|(0,0,1,0),\nonumber\\
&{\bm q}^+_{12}=|q^+_{12}|(1,2,1,1),\quad
{\bm q}^+_{13}=|q^+_{13}|(1,1,1,1),\quad
{\bm q}^+_{14}=|q^+_{14}|(1,1,0,1),\nonumber\\
&{\bm q}^+_{23}=|q^+_{23}|(0,1,1,1),\quad
{\bm q}^+_{24}=|q^+_{24}|(0,1,0,1),\quad
{\bm q}^+_{34}=|q^+_{34}|(0,0,0,1).
\label{d4zonoq}
\end{align}
For example, if we set $|q^+_{12}|=0$ for the coordinates ${\bm M}=(M_1,M_2,M_3,M_4)$ of the vertices in the parameter space of relative ranks, some of them change by $-{\bm q}^+_{12}$, while others remain.
We regard the zonotope generated by these vectors as the ${\cal Z}$ description.
Although the S-rule is not fully established for the $\widehat D$ quivers, the results in \eqref{d4zonoq} indicate that we can generalize the S-rule by allowing at most $|q^+_{ij}|$ D3-branes to connect the 5-brane $i$ and the mirror image of $j$ (see figure \ref{SforD}).

We can introduce an alternating notation
\begin{align}
&{\bm q}^-_{12}={\bm k}_1,\quad
{\bm q}^-_{13}={\bm k}_{12},\quad
{\bm q}^-_{14}={\bm k}_{123},\quad
{\bm q}^-_{23}={\bm k}_2,\quad
{\bm q}^-_{24}={\bm k}_{23},\quad
{\bm q}^-_{34}={\bm k}_3,\nonumber\\
&{\bm q}^+_{12}={\bm k}_{12234},\quad
{\bm q}^+_{13}={\bm k}_{1234},\quad
{\bm q}^+_{14}={\bm k}_{124},\quad
{\bm q}^+_{23}={\bm k}_{234},\quad
{\bm q}^+_{24}={\bm k}_{24},\quad
{\bm q}^+_{34}={\bm k}_4.
\end{align}
The notation is motivated by how the quantities $q^\mp_{ij}$ are constructed as linear combinations from the levels $k_1=q_1-q_2$, $k_2=q_2-q_3$, $k_3=q_3-q_4$ and $k_4=q_3+q_4$ in \eqref{kq}.
Namely, for example, since $q^+_{12}=q_1+q_2$ is constructed from the levels as $q^+_{12}=k_1+2k_2+k_3+k_4$, we denote the vector ${\bm q}^+_{12}$ as ${\bm k}_{12234}$ with the subscript $2$ doubled because of the non-trivial coefficient for $k_2$.
Interestingly, these numbers coincide with the directions $(1,2,1,1)$ appearing for ${\bm q}^+_{12}$ in \eqref{d4zonoq}.
As in the case of the $\widehat A$ quivers, it is interesting to note that the vectors ${\bm q}_{ij}^\mp$ correspond to the positive roots of the $\widehat D_4$ algebra.

We argue that, as in the previous case of the $\widehat A$ quivers, the three descriptions are equivalent.
Our argument is almost the same as previously.
First, we note that, as the S-rule does not depend on 5-brane orders for the $\widehat A$ quivers, the S-rule for the $\widehat D_4$ quiver is valid for all of the 192 transformations in the Weyl group.
Namely, once a configuration with relative ranks ${\bm M}$ is expressed by
\begin{align}
{\bm M}=\sum c_{ij}^\mp{\bm q}_{ij}^\mp,\quad 0\le c_{ij}^\mp\le 1,
\end{align}
with a certain choice of coefficients $c_{ij}^\mp$, another configuration obtained after applying one of the Weyl reflections $r_i$ with ${\bm M}'=r_i({\bm M})$ is also expressed by
\begin{align}
{\bm M}'=\sum c_{ij}^\mp{}'{\bm q}_{ij}^\mp{},\quad 0\le c_{ij}^\mp{}'\le 1,
\end{align}
with another choice of coefficients $c_{ij}^\mp{}'$.
For example, for the application of the Weyl reflection $r_4$, using the transformations in \eqref{d4weyl}, we find
\begin{align}
{\bm M}'=r_4({\bm M})
=\sum c_{ij}^\mp r^\text{H}_4({\bm q}_{ij}^\mp)+|q_{34}^+|(0,0,0,1),
\end{align}
where the transformation $r^\text{H}_i$ is defined from the homogeneous part of \eqref{d4weyl} and the transformations on ${\bm q}_{ij}^\mp$ can be found in table \ref{rqmp}.
Then, it is not difficult to find that by choosing the coefficients
\begin{align}
&c_{12}^-{}'=c_{12}^-,\quad
c_{13}^-{}'=c_{14}^+,\quad
c_{14}^-{}'=c_{13}^+,\quad
c_{23}^-{}'=c_{24}^+,\quad
c_{24}^-{}'=c_{23}^+,\quad
c_{34}^-{}'=c_{34}^-,\nonumber\\
&c_{12}^+{}'=c_{12}^+,\quad
c_{13}^+{}'=c_{14}^-,\quad
c_{14}^+{}'=c_{13}^-,\quad
c_{23}^+{}'=c_{24}^-,\quad
c_{24}^+{}'=c_{23}^-,\quad
c_{34}^+{}'=1-c_{34}^+,
\end{align}
the coefficients of ${\bm M}'$ satisfy $0\le c_{ij}^\mp{}'\le 1$.
The invariance also works for the other Weyl reflections $r_i$ which generate the Weyl group totally.

\begin{table}[t!]
\centering
\begin{tabular}{c|c|cccccccccccc}
&$(q_1,q_2,q_3,q_4)$&
${\bm q}^-_{12}$&${\bm q}^-_{13}$&${\bm q}^-_{14}$&
${\bm q}^-_{23}$&${\bm q}^-_{24}$&${\bm q}^-_{34}$&
${\bm q}^+_{12}$&${\bm q}^+_{13}$&${\bm q}^+_{14}$&
${\bm q}^+_{23}$&${\bm q}^+_{24}$&${\bm q}^+_{34}$\\\hline
$r^\text{H}_1$&$q_1\leftrightarrow q_2$
&
$-{\bm q}^-_{12}$&${\bm q}^-_{23}$&${\bm q}^-_{24}$&
${\bm q}^-_{13}$&${\bm q}^-_{14}$&${\bm q}^-_{34}$&
${\bm q}^+_{12}$&${\bm q}^+_{23}$&${\bm q}^+_{24}$&
${\bm q}^+_{13}$&${\bm q}^+_{14}$&${\bm q}^+_{34}$\\
$r^\text{H}_2$&$q_2\leftrightarrow q_3$
&
${\bm q}^-_{13}$&${\bm q}^-_{12}$&${\bm q}^-_{14}$&
$-{\bm q}^-_{23}$&${\bm q}^-_{34}$&${\bm q}^-_{24}$&
${\bm q}^+_{13}$&${\bm q}^+_{12}$&${\bm q}^+_{14}$&
${\bm q}^+_{23}$&${\bm q}^+_{34}$&${\bm q}^+_{24}$\\
$r^\text{H}_3$&$q_3\leftrightarrow q_4$
&
${\bm q}^-_{12}$&${\bm q}^-_{14}$&${\bm q}^-_{13}$&
${\bm q}^-_{24}$&${\bm q}^-_{23}$&$-{\bm q}^-_{34}$&
${\bm q}^+_{12}$&${\bm q}^+_{14}$&${\bm q}^+_{13}$&
${\bm q}^+_{24}$&${\bm q}^+_{23}$&${\bm q}^+_{34}$\\
$r^\text{H}_4$&$q_3\leftrightarrow -q_4$
&
${\bm q}^-_{12}$&${\bm q}^+_{14}$&${\bm q}^+_{13}$&
${\bm q}^+_{24}$&${\bm q}^+_{23}$&${\bm q}^-_{34}$&
${\bm q}^+_{12}$&${\bm q}^-_{14}$&${\bm q}^-_{13}$&
${\bm q}^-_{24}$&${\bm q}^-_{23}$&$-{\bm q}^+_{34}$
\end{tabular}
\caption{Homogeneous transformations of Weyl reflections $r^\text{H}_i$ on ${\bm q}_{ij}^\mp$.}
\label{rqmp}
\end{table}

Then, as in the $\widehat A$ quivers, it is clear that ${\cal V}\subset{\cal Z}\subset{\cal H}$, since the ${\cal Z}$ description is invariant under the Weyl reflections.
Each configuration with levels and ranks also induces a cone in the parameter space of relative ranks.
By the same argument as the $\widehat A$ quivers, we regard the three descriptions as identical.

We can also repeat the arguments of the relation among the facets.
As in \eqref{facetsumAn}, we encounter the facet of the truncated octahedron ($A_3$) with $4!=24$ vertices by removing the nodes \textcircled{\scriptsize 1}, \textcircled{\scriptsize 3}, \textcircled{\scriptsize 4} (and the affine node \textcircled{\scriptsize 0}) from the $\widehat D_4$ quiver, while we encounter the facet of the parallelepiped ($(A_1)^3$) with $(2!)^3=8$ vertices by removing the non-affine node \textcircled{\scriptsize 2}.
Each of the former case gives $192/24=8$ facets while the latter case gives $192/8=24$ facets.
This estimation matches $8\times 3=24$ inequalities in \eqref{d4affineineq} and $24$ inequalities in \eqref{d4nonaffineineq}.

\subsubsection{Discrete translations}

\begin{table}[t!]
\centering
\begin{tabular}{c|c}
inequalities&translations $\Delta{\bm M}$\\\hline
$M_1\ge 0$&
$\Delta_1{\bm M}$\\
$-M_1+|q^-_{12}|+|q^+_{12}|+|q^-_{13}|+|q^+_{13}|+|q^-_{14}|+|q^+_{14}|\ge 0$&
$-\Delta_1{\bm M}$\\
$-M_1+M_2+|q^-_{12}|\ge 0$&
$-\Delta_1{\bm M}+\Delta_2{\bm M}$\\
$M_1-M_2+|q^+_{12}|+|q^-_{23}|+|q^+_{23}|+|q^-_{24}|+|q^+_{24}|\ge 0$&
$\Delta_1{\bm M}-\Delta_2{\bm M}$\\
$-M_2+M_3+M_4+|q^-_{13}|+|q^-_{23}|\ge 0$&
$-\Delta_2{\bm M}+\Delta_3{\bm M}+\Delta_4{\bm M}$\\
$M_2-M_3-M_4+|q^+_{13}|+|q^+_{23}|+|q^-_{34}|+|q^+_{34}|\ge 0$&$\Delta_2{\bm M}-\Delta_3{\bm M}-\Delta_4{\bm M}$\\
$M_3-M_4+|q^+_{14}|+|q^+_{24}|+|q^+_{34}|\ge 0$&$\Delta_3{\bm M}-\Delta_4{\bm M}$\\
$-M_3+M_4+|q^-_{14}|+|q^-_{24}|+|q^-_{34}|\ge 0$&$-\Delta_3{\bm M}+\Delta_4{\bm M}$\\\hline
$M_3\ge 0$&$\Delta_3{\bm M}$\\
$-M_3+|q^+_{12}|+|q^+_{13}|+|q^+_{23}|+|q^-_{14}|+|q^-_{24}|+|q^-_{34}|\ge 0$&$-\Delta_3{\bm M}$\\
$M_2-M_3+|q^-_{34}|\ge 0$&$\Delta_2{\bm M}-\Delta_3{\bm M}$\\
$-M_2+M_3+|q^+_{12}|+|q^-_{13}|+|q^-_{23}|+|q^+_{14}|+|q^+_{24}|\ge 0$&$-\Delta_2{\bm M}+\Delta_3{\bm M}$\\
$M_1-M_2+M_4+|q^-_{23}|+|q^-_{24}|\ge 0$&$\Delta_1{\bm M}-\Delta_2{\bm M}+\Delta_4{\bm M}$\\
$-M_1+M_2-M_4+|q^-_{12}|+|q^+_{13}|+|q^+_{14}|+|q^+_{34}|\ge 0$&$-\Delta_1{\bm M}+\Delta_2{\bm M}-\Delta_4{\bm M}$\\
$-M_1+M_4+|q^-_{12}|+|q^-_{13}|+|q^-_{14}|\ge 0$&$-\Delta_1{\bm M}+\Delta_4{\bm M}$\\
$M_1-M_4+|q^+_{23}|+|q^+_{24}|+|q^+_{34}|\ge 0$&$\Delta_1{\bm M}-\Delta_4{\bm M}$\\\hline
$M_4\ge 0$&
$\Delta_4{\bm M}$\\
$-M_4+|q^+_{12}|+|q^+_{13}|+|q^+_{23}|+|q^+_{14}|+|q^+_{24}|+|q^+_{34}|\ge 0$&
$-\Delta_4{\bm M}$\\
$M_2-M_4+|q^+_{34}|\ge 0$&
$\Delta_2{\bm M}-\Delta_4{\bm M}$\\
$-M_2+M_4+|q^+_{12}|+|q^-_{13}|+|q^-_{23}|+|q^-_{14}|+|q^-_{24}|\ge 0$&
$-\Delta_2{\bm M}+\Delta_4{\bm M}$\\
$M_1-M_2+M_3+|q^-_{23}|+|q^+_{24}|\ge 0$&$\Delta_1{\bm M}-\Delta_2{\bm M}+\Delta_3{\bm M}$\\
$-M_1+M_2-M_3+|q^-_{12}|+|q^+_{13}|+|q^-_{14}|+|q^-_{34}|\ge 0$&
$-\Delta_1{\bm M}+\Delta_2{\bm M}-\Delta_3{\bm M}$\\
$M_1-M_3+|q^+_{23}|+|q^-_{24}|+|q^-_{34}|\ge 0$&
$\Delta_1{\bm M}-\Delta_3{\bm M}$\\
$-M_1+M_3+|q^-_{12}|+|q^-_{13}|+|q^+_{14}|\ge 0$&
$-\Delta_1{\bm M}+\Delta_3{\bm M}$
\end{tabular}
\caption{Inequalities obtained by requiring that the rank of the affine node \textcircled{\scriptsize 0} is lower than other ranks of the potentially affine nodes \textcircled{\scriptsize 1}, \textcircled{\scriptsize 3} and \textcircled{\scriptsize 4}.
When the inequality is violated, the relative rank ${\bm M}$ is transformed by the translation $\Delta{\bm M}$ in the parameter space of ${\bm M}$ through the outer automorphism and the subsequent Weyl reflections explained around \eqref{Mtransl1}.
Note that the coefficients of the translations $\Delta_i{\bm M}$ can be read off unambiguously from the integral coefficients of $M_i$ in the inequality.
Each block of inequalities corresponds respectively to the nodes \textcircled{\scriptsize 1}, \textcircled{\scriptsize 3} and \textcircled{\scriptsize 4}.}
\label{D4affine}
\end{table}

After clarifying the three descriptions, let us start our general arguments for the transformations of relative ranks under duality cascades as in the previous sample calculation in figure \ref{d4dualitycascade}.
We shall see that, as in the $\widehat A$ quivers, the transformations are compatible discrete translations in the space of relative ranks.
This allows us to answer the questions on the finiteness and the uniqueness.

Suppose that one of the ranks, say $N_1(=N+M_1)$, is lower than the reference rank $N_0(=N)$, $M_1<0$.
Then, we change references by applying the outer automorphism $\sigma_1$ and redefine the reference rank by $N'=N+M_1$,
\begin{align}
\begin{matrix}N'-M_1\\N'\end{matrix}\quad 2N'-2M_1+M_2\quad\begin{matrix}N'-M_1+M_4\\N'-M_1+M_3\end{matrix}
\end{align}
where only ranks in the $\widehat D_4$ quiver in figure \ref{d4} are shown.
Since we have applied the outer automorphism $\sigma_1$, we have changed the levels $(k_0,k_1,k_2,k_3,k_4)$ into $(k_1,k_0,k_2,k_4,k_3)$ which is equivalent to the change of $(q_1,q_2,q_3,q_4)$ into $(-q_1,q_2,q_3,-q_4)$.
To return to the standard order, we need to apply a series of Weyl reflections $r_1r_2r_4r_3r_2r_1$.
After some rather tedious but straightforward computations (if done by hand), finally, we arrive at
\begin{align}
\begin{matrix}N'+M'_1\\N'\end{matrix}\quad
2N'+M'_2\quad\begin{matrix}N'+M'_3\\N'+M'_4\end{matrix}
\label{Mtransl1}
\end{align}
with the overall rank $N$ and the relative ranks ${\bm M}=(M_1,M_2,M_3,M_4)$ transformed as $N'=N+M_1$ and ${\bm M}'={\bm M}+\Delta_1{\bm M}$, with $\Delta_1{\bm M}$ defined later in \eqref{DMq}.
Note that, as in the previous case for the $\widehat A$ quivers (which is guaranteed by the charge conservation $Q_\text{RR}$), the transformation can be regarded as a discrete translation in the parameter space of relative ranks ${\bm M}=(M_1,M_2,M_3,M_4)$ in the direction where the breakdown of the inequality $M_1<0$ is alleviated.
This is the first crucial sign that the working hypothesis of duality cascades works consistently for the $\widehat D_4$ quiver.

In addition to the case of $M_1<0$, we can similarly apply to other inequalities and find
\begin{align}
\Delta_1{\bm M}&={\bm q}^-_{12}+{\bm q}^-_{13}+{\bm q}^-_{14}+{\bm q}^+_{12}+{\bm q}^+_{13}+{\bm q}^+_{14},\nonumber\\
\Delta_2{\bm M}&={\bm q}^-_{13}+{\bm q}^-_{14}+{\bm q}^-_{23}+{\bm q}^-_{24}+2{\bm q}^+_{12}+{\bm q}^+_{13}+{\bm q}^+_{14}+{\bm q}^+_{23}+{\bm q}^+_{24},\nonumber\\
\Delta_3{\bm M}&={\bm q}^-_{14}+{\bm q}^-_{24}+{\bm q}^-_{34}+{\bm q}^+_{12}+{\bm q}^+_{13}+{\bm q}^+_{23},\nonumber\\
\Delta_4{\bm M}&={\bm q}^+_{12}+{\bm q}^+_{13}+{\bm q}^+_{14}+{\bm q}^+_{23}+{\bm q}^+_{24}+{\bm q}^+_{34}.
\label{DMq}
\end{align}
For example, for the cases of $M_3<0$ and $M_4<0$, duality cascades work similarly and the relative ranks change by the discrete translations $\Delta_3{\bm M}$ and $\Delta_4{\bm M}$ respectively.
Moreover, since the variable $M_2$ also appears in the potentially affine nodes in the application of Weyl reflections \eqref{d4weyl}, we encounter $\Delta_2{\bm M}$.
However, as we have noted in figure \ref{d4dualitycascade} and clarify more below around \eqref{d4ineqM2}, the inequality $M_2\ge 0$ does not induce discrete translations as there are no outer automorphisms relating the non-affine node \textcircled{\scriptsize 2} with the affine node \textcircled{\scriptsize 0}.
For all the inequalities in \eqref{d4affineineq}, the discrete translations are given in table \ref{D4affine}.
As the second crucial point, it is interesting to note that the discrete translations are given only by four vectors with integral coefficients, whose number matches the dimension of the parameter space of relative ranks.
This implies that the polytope extends to the parameter space through compatible discrete translations.

The results on discrete translations \eqref{DMq} are more impressive if given in the basis of ${\bm k}$,
\begin{align}
\Delta_1{\bm M}&={\bm k}_1+{\bm k}_{12}+{\bm k}_{123}+{\bm k}_{12234}+{\bm k}_{1234}+{\bm k}_{124},\nonumber\\
\Delta_2{\bm M}&={\bm k}_{12}+{\bm k}_{123}+{\bm k}_2+{\bm k}_{23}+2{\bm k}_{12234}+{\bm k}_{1234}+{\bm k}_{124}+{\bm k}_{234}+{\bm k}_{24},\nonumber\\
\Delta_3{\bm M}&={\bm k}_{123}+{\bm k}_{23}+{\bm k}_3+{\bm k}_{12234}+{\bm k}_{1234}+{\bm k}_{234},\nonumber\\
\Delta_4{\bm M}&={\bm k}_{12234}+{\bm k}_{1234}+{\bm k}_{124}+{\bm k}_{234}+{\bm k}_{24}+{\bm k}_4.
\label{DMk}
\end{align}
In this expression, it is interesting to observe that ${\bm k}_{\cdots i^{n_i}\cdots}$ (with $n_i$ denoting the multiplicity of $i$) appears in $\Delta_i{\bm M}$ with coefficients $n_i$.
For example, ${\bm k}_{12234}={\bm k}_{12^234}$ appear in $\Delta_2{\bm M}$ with a coefficient $2$.

\begin{table}[!t]
\centering
\begin{tabular}{c|c}
inequalities&mis.\\\hline
$M_2\ge 0$&${\bm q}^+_{12}$\\
$-M_2+2|q^+_{12}|+|q^-_{13}|+|q^-_{23}|+|q^+_{13}|+|q^+_{23}|+|q^-_{14}|+|q^-_{24}|+|q^+_{14}|+|q^+_{24}|\ge 0$&\\
$2M_1-M_2+|q^-_{23}|+|q^+_{23}|+|q^-_{24}|+|q^+_{24}|\ge 0$&${\bm q}^-_{12}$\\
$-2M_1+M_2+2|q^-_{12}|+|q^-_{13}|+|q^+_{13}|+|q^-_{14}|+|q^+_{14}|\ge 0$&\\
$2M_3-M_2+|q^-_{13}|+|q^-_{23}|+|q^+_{14}|+|q^+_{24}|\ge 0$&${\bm q}^-_{34}$\\
$-2M_3+M_2+|q^+_{13}|+|q^+_{23}|+|q^-_{14}|+|q^-_{24}|+2|q^-_{34}|\ge 0$&\\
$2M_4-M_2+|q^-_{13}|+|q^-_{23}|+|q^-_{14}|+|q^-_{24}|\ge 0$&${\bm q}^+_{34}$\\
$-2M_4+M_2+|q^+_{13}|+|q^+_{23}|+|q^+_{14}|+|q^+_{24}|+2|q^+_{34}|\ge 0$&\\\hline
$-M_1+2M_2-M_3-M_4+|q^-_{12}|+|q^+_{13}|+|q^-_{34}|+|q^+_{34}|\ge 0$&${\bm q}^-_{23}$\\
$M_1-2M_2+M_3+M_4+|q^+_{12}|+|q^-_{13}|+2|q^-_{23}|+|q^-_{24}|+|q^+_{24}|\ge 0$&\\
$-M_1+M_3+M_4+|q^-_{12}|+|q^-_{13}|\ge 0$&${\bm q}^+_{23}$\\
$M_1-M_3-M_4+|q^+_{12}|+|q^+_{13}|+2|q^+_{23}|+|q^-_{24}|+|q^-_{34}|+|q^+_{24}|+|q^+_{34}|\ge 0$&\\
$M_1-M_3+M_4+|q^-_{24}|+|q^-_{34}|\ge 0$&${\bm q}^+_{14}$\\
$-M_1+M_3-M_4+|q^-_{12}|+|q^+_{12}|+|q^-_{13}|+|q^+_{13}|+2|q^+_{14}|+|q^+_{24}|+|q^+_{34}|\ge 0$&\\
$M_1+M_3-M_4+|q^+_{24}|+|q^+_{34}|\ge 0$&${\bm q}^-_{14}$\\
$-M_1-M_3+M_4+|q^-_{12}|+|q^+_{12}|+|q^-_{13}|+|q^+_{13}|+2|q^-_{14}|+|q^-_{24}|+|q^-_{34}|\ge 0$&\\\hline
$M_1-M_2+M_3+M_4+|q^-_{23}|\ge 0$&${\bm q}^+_{13}$\\
$-M_1+M_2-M_3-M_4+|q^-_{12}|+|q^+_{12}|+2|q^+_{13}|+|q^+_{23}|+|q^-_{14}|+|q^-_{34}|+|q^+_{14}|+|q^+_{34}|\ge 0$&\\
$M_1+M_2-M_3-M_4+|q^+_{23}|+|q^-_{34}|+|q^+_{34}|\ge 0$&${\bm q}^-_{13}$\\
$-M_1-M_2+M_3+M_4+|q^-_{12}|+|q^+_{12}|+2|q^-_{13}|+|q^-_{23}|+|q^-_{14}|+|q^+_{14}|\ge 0$&\\
$-M_1+M_2+M_3-M_4+|q^-_{12}|+|q^+_{14}|+|q^+_{34}|\ge 0$&${\bm q}^-_{24}$\\
$M_1-M_2-M_3+M_4+|q^+_{12}|+|q^-_{23}|+|q^+_{23}|+|q^-_{14}|+2|q^-_{24}|+|q^-_{34}|\ge 0$&\\
$-M_1+M_2-M_3+M_4+|q^-_{12}|+|q^-_{14}|+|q^-_{34}|\ge 0$&${\bm q}^+_{24}$\\
$M_1-M_2+M_3-M_4+|q^+_{12}|+|q^-_{23}|+|q^+_{23}|+|q^+_{14}|+2|q^+_{24}|+|q^+_{34}|\ge 0$&
\end{tabular}
\caption{Inequalities obtained by comparing the rank of the affine node \textcircled{\scriptsize 0} with the node \textcircled{\scriptsize 2}, which will never be the affine node through the outer automorphism.
The vertices on the two opposite parallel facets are distinct by the translation expected from table \ref{D4affine} with a small misalignment.
The other misalignments in the same block are identical to the vectors generating the parallelepiped $(A_1)^3$ on the facet.
Totally the four misalignments in the same block form a gap in tiling the space.
For example, both of the facets on ${\pm M_2}+\cdots=0$ are the same parallelepipeds generated by ${\bm q}_{12}^-$, ${\bm q}_{34}^-$, ${\bm q}_{34}^+$ whose vertices are separated by $\Delta_2{\bm M}-{\bm q}_{12}^+$.}
\label{D4nonaffine}
\end{table}

The other 24 inequalities \eqref{d4nonaffineineq} originate from comparing the rank of the affine node \textcircled{\scriptsize 0} with that of the non-affine one \textcircled{\scriptsize 2}.
Let us investigate them to understand the polytope more extensively.
Since there are two parallel facets
\begin{align}
M_2\ge 0,\quad
-M_2+2|q^+_{12}|+|q^-_{13}|+|q^-_{23}|+|q^+_{13}|+|q^+_{23}|+|q^-_{14}|+|q^-_{24}|+|q^+_{14}|+|q^+_{24}|\ge 0,
\label{d4ineqM2}
\end{align}
for the polytope to be a parallelotope, we may expect that the corresponding vertices on the two parallel facets are different by $\Delta_2{\bm M}$, even though the above arguments of discrete translations (through the applications of the outer automorphism and the Weyl reflections when one of the inequalities is violated) does not work for the non-affine node \textcircled{\scriptsize 2}.
This is, however, not the case.
Instead, we find that the distance is
\begin{align}
{\bm k}_{12}+{\bm k}_{123}+{\bm k}_2+{\bm k}_{23}+{\bm k}_{12234}+{\bm k}_{1234}+{\bm k}_{124}+{\bm k}_{234}+{\bm k}_{24},
\end{align}
shortened by ${\bm k}_{12234}={\bm q}^+_{12}$ compared with $\Delta_2{\bm M}$ in \eqref{DMk}, where the coefficient of ${\bm k}_{12234}$ is the non-trivial mark $2$.
Namely, the polytope still tiles the entire space without overlaps, though there appear some gaps.
Since the misalignment occurs for the node with the non-trivial mark, we may attribute the misalignment to the non-trivial mark in the Dynkin diagram of the $\widehat D_4$ quiver.
So far we have only consider two inequalities in \eqref{d4ineqM2}.
The other inequalities are studied in table \ref{D4nonaffine}.

If we require the volume of the gap to vanish, we find the condition $q^-_{12}q^-_{34}q^+_{34}\times q^+_{12}=0$ since the facet is a parallelepiped $(A_1)^3$ with the edge lengths being $q^-_{12}$, $q^-_{34}$, $q^+_{34}$ and the misalignment is $q^+_{12}$.
After taking care of the other conditions of no gaps, the total condition of no gaps is
\begin{align}
q^-_{12}q^-_{34}q^+_{12}q^+_{34}=0,\quad
q^-_{13}q^-_{24}q^+_{13}q^+_{24}=0,\quad
q^-_{14}q^-_{23}q^+_{14}q^+_{23}=0.
\end{align}
The condition is solved when three of $q_i^2$ are equal.
It is interesting to note that the level assignment $(-k,0,0,0,k)$ studied in \cite{MN4,KN} satisfies this condition.
This may suggest that the reason that this level assignment is tractable is because the gaps are filled.

We have studied duality cascades for the $\widehat D_4$ quiver in this subsection.
The study is possible by adopting the working hypothesis of duality cascades given in the group-theoretical language.
Originally, it is unclear whether duality cascades always terminate in finite steps and whether the final destination is uniquely determined by an initial configuration regardless of processes of duality cascades.
These questions are rephrased geometrically into the question whether the associated polytope tiles the parameter space by discrete translations.
We have found a partially positive answer stating that all the discrete translations are still compatible in the parameter space of relative ranks.
However, since duality cascades can only be activated for the potentially affine nodes, when we compare the non-affine node with the affine node, this introduces some gaps in tiling.
The gaps are filled for the special choice of levels $(-k,0,0,0,k)$ adopted in \cite{MN4,KN}.
This implies that, with large ranks for the non-affine node, duality cascades still terminate uniquely.

\subsection{$\widehat D_5$ quiver}

\begin{figure}[t!]
\centering\includegraphics[scale=0.5,angle=-90]{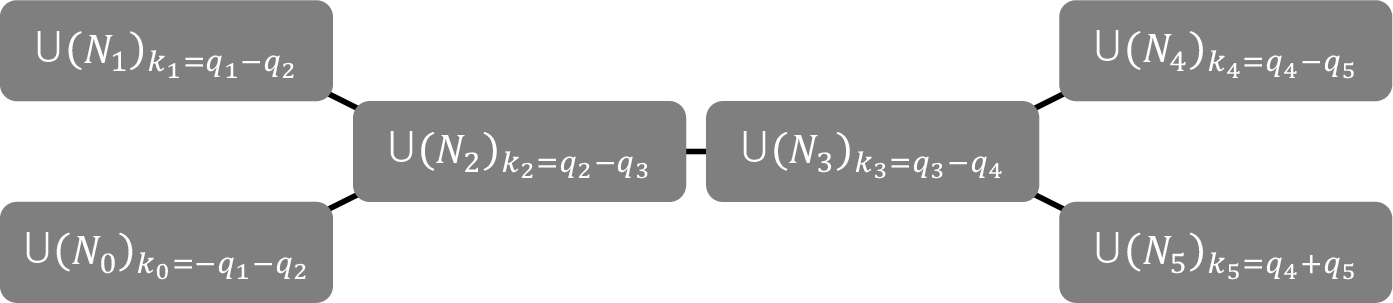}
\caption{$\widehat D_5$ quiver.}
\label{d5}
\end{figure}

Now let us turn to the super Chern-Simons theory of the $\widehat D_5$ quiver (see figure \ref{d5}).
The gauge group and the bifundamental matters are read off directly from the quiver diagram as in the previous case of the $\widehat D_4$ quiver in figure \ref{d4}.
It is important to note that the outer automorphism for the $\widehat D_5$ quiver is not ${\mathbb Z}_2\times{\mathbb Z}_2$ but ${\mathbb Z}_4$,
\begin{align}
\sigma&:k_0\to k_4,\;k_1\to k_5,\;k_2\to k_3,\;k_3\to k_2,\;k_4\to k_1,\;k_5\to k_0,\nonumber\\
&\quad N_0\to N_4,\;N_1\to N_5,\;N_2\to N_3,\;N_3\to N_2,\;N_4\to N_1,\;N_5\to N_0.
\end{align}
Other than the outer automorphism, most of the studies in the previous subsection work similarly.
The duality transformations or the Weyl reflections can be inferred easily as in \eqref{d4weyl} and we abbreviate them here.
As before, we label the ranks as
\begin{align}
N_0=N,\;
N_1=N+M_1,\;
N_2=2N+M_2,\;
N_3=2N+M_3,\;
N_4=N+M_4,\;
N_5=N+M_5,
\end{align}
with the marks included for the coefficients of the ``overall'' rank $N$ and we denote the levels as
\begin{align}
k_0=-q_1-q_2,\;
k_1=q_1-q_2,\;
k_2=q_2-q_3,\;
k_3=q_3-q_4,\;
k_4=q_4-q_5,\;
k_5=q_4+q_5.
\end{align}
With the Weyl reflections and the outer automorphism, we can adopt the working hypothesis and study examples of duality cascades.
Then, we encounter the same questions on the finiteness and the uniqueness.

To answer these questions, as in the case of the $\widehat D_4$ quiver, we can define the polytope associated to duality cascades for the $\widehat D_5$ quiver with the three descriptions, ${\cal H}$, ${\cal Z}$ and ${\cal V}$.
As in \eqref{d4affineineq}, \eqref{d4nonaffineineq}, for the ${\cal H}$ description, we write down the inequalities by applying the Weyl reflections to the configurations and require the rank of the affine node to stay the lowest.
For the ${\cal V}$ description, we obtain various configurations of levels without relative ranks using the Weyl reflections and return to the original level to read off the relative ranks.
The ${\cal Z}$ description is obtained from the ${\cal V}$ description by setting one of the variables $q^\mp_{ij}=q_i\mp q_j$ to zero.
After performing these computations, as a result we find that the polytope has 162 facets and 1920 vertices.
As before the number of vertices matches the order of the $D_5$ Weyl group $\#(W(D_5))=1920$.
In the ${\cal Z}$ description, it is a zonotope generated by 20 vectors
\begin{align}
&{\bm q}^-_{12}={\bm k}_{1},\;
{\bm q}^-_{13}={\bm k}_{12},\;
{\bm q}^-_{14}={\bm k}_{123},\;
{\bm q}^-_{15}={\bm k}_{1234},\nonumber\\
&{\bm q}^-_{23}={\bm k}_{2},\;
{\bm q}^-_{24}={\bm k}_{23},\;
{\bm q}^-_{25}={\bm k}_{234},\;
{\bm q}^-_{34}={\bm k}_{3},\;
{\bm q}^-_{35}={\bm k}_{34},\;
{\bm q}^-_{45}={\bm k}_{4},\nonumber\\
&{\bm q}^+_{12}={\bm k}_{1223345},\;
{\bm q}^+_{13}={\bm k}_{123345},\;
{\bm q}^+_{14}={\bm k}_{12345},\;
{\bm q}^+_{15}={\bm k}_{1235},\nonumber\\
&{\bm q}^+_{23}={\bm k}_{23345},\;
{\bm q}^+_{24}={\bm k}_{2345},\;
{\bm q}^+_{25}={\bm k}_{235},\;
{\bm q}^+_{34}={\bm k}_{345},\;
{\bm q}^+_{35}={\bm k}_{35},\;
{\bm q}^+_{45}={\bm k}_{5},
\end{align}
with the two expressions explicitly given by, for example, ${\bm q}^+_{12}={\bm k}_{1223345}=|q^+_{12}|(1,2,2,1,1)=|k_1+2k_2+2k_3+k_4+k_5|(1,2,2,1,1)$ in the space of relative ranks ${\bm M}=(M_1,M_2,M_3,M_4,M_5)$.
These vectors correspond to the positive roots of the $D_5$ algebra.

As \eqref{Mtransl1} in the case of the $\widehat D_4$ quiver, when the inequalities originating from the potentially affine node are violated, duality cascades induce the discrete translations so that the violation is alleviated through the outer automorphism and the subsequent Weyl reflections.
After similar computations, finally we find that the discrete translations are generated by
\begin{align}
\Delta_1{\bm M}&={\bm q}^-_{12}+{\bm q}^-_{13}+{\bm q}^-_{14}+{\bm q}^-_{15}+{\bm q}^+_{12}+{\bm q}^+_{13}+{\bm q}^+_{14}+{\bm q}^+_{15},\nonumber\\
\Delta_2{\bm M}&={\bm q}^-_{13}+{\bm q}^-_{14}+{\bm q}^-_{15}+{\bm q}^-_{23}+{\bm q}^-_{24}+{\bm q}^-_{25}+2{\bm q}^+_{12}+{\bm q}^+_{13}+{\bm q}^+_{14}+{\bm q}^+_{15}+{\bm q}^+_{23}+{\bm q}^+_{24}+{\bm q}^+_{25},\nonumber\\
\Delta_3{\bm M}&={\bm q}^-_{14}+{\bm q}^-_{15}+{\bm q}^-_{24}+{\bm q}^-_{25}+{\bm q}^-_{34}+{\bm q}^-_{35}\nonumber\\
&\qquad\qquad\qquad+2{\bm q}^+_{12}+2{\bm q}^+_{13}+{\bm q}^+_{14}+{\bm q}^+_{15}+2{\bm q}^+_{23}+{\bm q}^+_{24}+{\bm q}^+_{25}+{\bm q}^+_{34}+{\bm q}^+_{35},\nonumber\\
\Delta_4{\bm M}&={\bm q}^-_{15}+{\bm q}^-_{25}+{\bm q}^-_{35}+{\bm q}^-_{45}+{\bm q}^+_{12}+{\bm q}^+_{13}+{\bm q}^+_{14}+{\bm q}^+_{23}+{\bm q}^+_{24}+{\bm q}^+_{34},\nonumber\\
\Delta_5{\bm M}&={\bm q}^+_{12}+{\bm q}^+_{13}+{\bm q}^+_{14}+{\bm q}^+_{15}+{\bm q}^+_{23}+{\bm q}^+_{24}+{\bm q}^+_{25}+{\bm q}^+_{34}+{\bm q}^+_{35}+{\bm q}^+_{45}.
\label{d5DeltaM}
\end{align}
It is interesting to note that the discrete translations are given only by five independent vectors with integral coefficients in the five-dimensional space.
Also, the distance between the vertices on the two facets $\pm M_2+\cdots=0$ (or $\pm M_3+\cdots=0$) is shortened from $\Delta_2{\bm M}$ (or $\Delta_3{\bm M}$ respectively) in \eqref{d5DeltaM} by the vectors with the non-trivial mark 2.
For this reason, the polytope tiles the whole parameter space of relative ranks but with some gaps again.

\subsection{$\widehat D_n$ quiver}

\begin{figure}[t!]
\centering\includegraphics[scale=0.5,angle=-90]{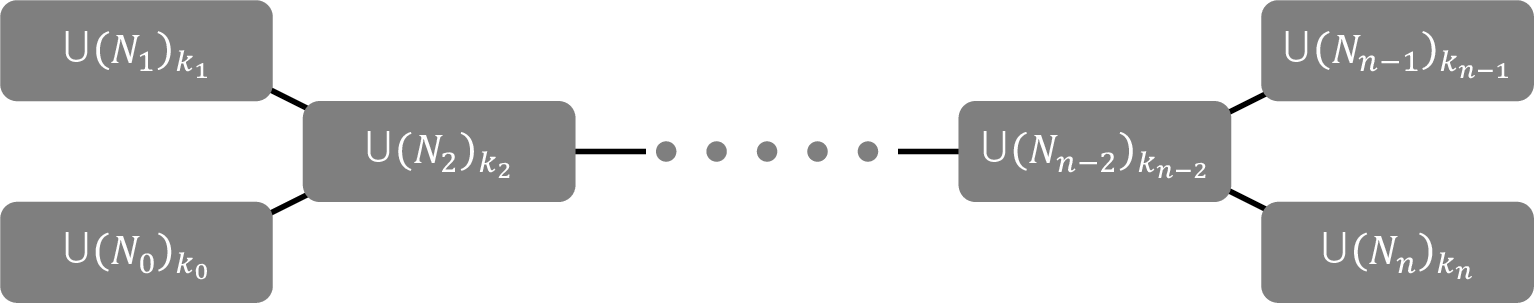}
\caption{$\widehat D_n$ quiver.}
\label{dn}
\end{figure}

We can repeat the same analysis for general $\widehat D_n$ quivers (see figure \ref{dn}).
It is important to note that the outer automorphism is ${\mathbb Z}_2\times{\mathbb Z}_2$ for even $n$, while it is ${\mathbb Z}_4$ for odd $n$.
Other than that, most of the analysis is parallel.

Here let us provide an evaluation of the number of facets.
By removing the node \textcircled{\scriptsize $j$} ($1\le j\le n-3$) in the $D_n$ Dynkin diagram, we reach the subgroup of $A_{j-1}\times D_{n-j}$ which leads to $2^{n-1}n!/[j!\times 2^{n-j-1}(n-j)!]$ facets.
By removing the node \textcircled{\scriptsize $j$} with $j=n-2$, we reach the subgroup of $A_{n-3}\times A_1\times A_1$ which leads to $2^{n-1}n!/[(n-2)!\times 2\times 2]$ facets.
(This is the same expression as for $1\le j\le n-3$.)
By removing the node \textcircled{\scriptsize $j$} with $j=n-1$ or $j=n$, we reach the subgroup of $A_{n-1}$ which leads to $2^{n-1}n!/n!$ facets.
Totally, the number of facets is
\begin{align}
\sum_{j=1}^{n-2}\frac{2^{n-1}n!}{j!\times 2^{n-j-1}(n-j)!}+\frac{2^{n-1}n!}{n!}+\frac{2^{n-1}n!}{n!}
=3^n-n\cdot 2^{n-1}-1.
\end{align}
Interestingly, this coincides with the evaluation in studying generalizations of permutohedrons in \cite{JSD}.

\section{$\widehat E$ quivers}\label{E}

\begin{table}[!t]
\centering
\begin{tabular}{c|ccl}
quivers&vertices&zonotopes&facets\\\hline
$\widehat E_6$&$51840$&$36$&$(27,216,720,72,216,27)$\\
$\widehat E_7$&$2903040$&$63$&$(126,2016,10080,576,4032,756, 56)$\\
$\widehat E_8$&$696729600$&$120$&$(2160,69120,483840,17280,241920,60480,6720,240)$
\end{tabular}
\caption{Various geometric quantities of the polytope associated to duality cascades for the $\widehat E$ quivers.
We list the numbers of vertices, the numbers of vectors for zonotopes and the numbers of facets.
For the numbers of facets, we list separately for different types.
All of these quantities have a clear interpretation in the group theory.
For example, for the facet of the third type for the $\widehat E_6$ quiver, since we are left with $(A_2)^2\times A_1$ after removing the node \textcircled{\scriptsize 3} (and the affine node \textcircled{\scriptsize 0}) from the quiver in figure \ref{e6}, the number of facets is $51840/((3!)^2\times 2!)=720$.}
\label{egeometry}
\end{table}

Now let us turn to the super Chern-Simons theories of the $\widehat E$ quivers.
As in the previous cases, we can rewrite every step in duality cascades in terms of the group-theoretical language.
We first write down the duality transformations inspired by the Weyl reflections.
With the Weyl reflections and the outer automorphism, we adopt the working hypothesis of duality cascades.
Then, similarly, the questions on the finiteness and the uniqueness arise.
To answer them, we study the polytope associated to duality cascades in the three descriptions, ${\cal H}$, ${\cal Z}$ and ${\cal V}$.
Since the studies for the $\widehat E$ quivers are the same as those for the $\widehat D$ quivers, we simplify our presentation by mainly showing the results.

As previously, we find that various geometric properties of the polytope associated to duality cascades are also given by group-theoretical quantities.
For example, the number of vertices is given by the order of the Weyl groups and the number of vectors for the zonotope is given by the number of positive roots.
Also, the number of facets of each type is determined by the subgroup of the Weyl group which fixes the facet.
We list various geometric quantities in table \ref{egeometry} and see that these understandings work for $\widehat E$ as well in this section.
All of these studies imply that our questions on the finiteness and the uniqueness are answered positively under certain restrictions.

\subsection{$\widehat E_6$ quiver}

\begin{figure}[t!]
\centering\includegraphics[scale=0.5,angle=-90]{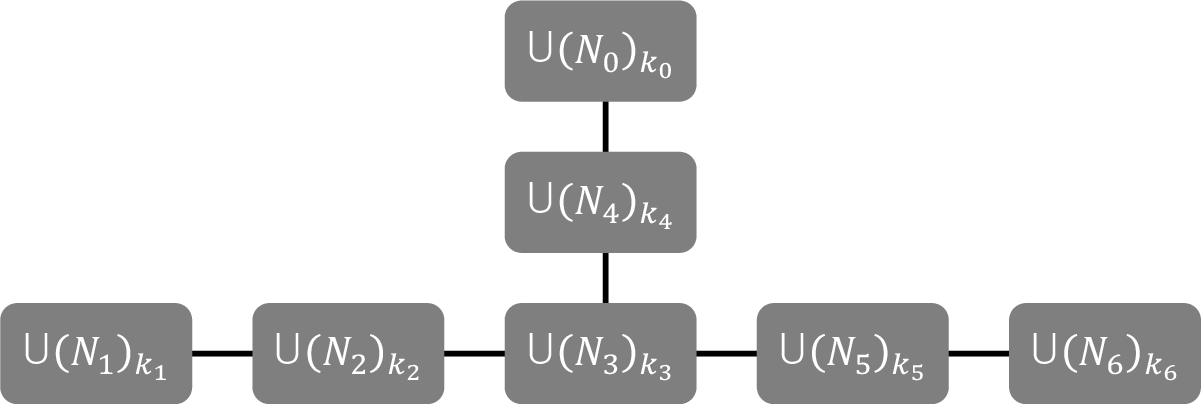}
\caption{$\widehat E_6$ quiver.}
\label{e6}
\end{figure}

Let us start with the $\widehat E_6$ quiver.
The levels and the ranks for the super Chern-Simons theory of the $\widehat E_6$ quiver are labeled as in figure \ref{e6} where levels are subject to the constraint $k_0+k_1+2k_2+3k_3+2k_4+2k_5+k_6=0$.
The duality transformations inspired by the Weyl reflections are given by
\begin{align}
r_1&: k_1\to-k_1,k_2\to k_2+k_1,N_1\to N_2-N_1+|k_1|,\nonumber\\
r_2&: k_2\to-k_2,k_1\to k_1+k_2,k_3\to k_3+k_2,N_2\to N_1+N_3-N_2+|k_2|,\nonumber\\
r_3&: k_3\to-k_3,k_2\to k_2+k_3,k_4\to k_4+k_3,k_5\to k_5+k_3,N_3\to N_2+N_4+N_5-N_3+|k_3|,\nonumber\\
r_4&: k_4\to-k_4,k_0\to k_0+k_4,k_3\to k_3+k_4,N_4\to N_3+N_0-N_4+|k_4|,\nonumber\\
r_5&: k_5\to-k_5,k_3\to k_3+k_5,k_6\to k_6+k_5,N_5\to N_3+N_6-N_5+|k_5|,\nonumber\\
r_6&: k_6\to-k_6,k_5\to k_5+k_6,N_6\to N_5-N_6+|k_6|,
\label{e6weyl}
\end{align}
and the outer automorphism ${\mathbb Z}_3$ is
\begin{align}
\sigma&:k_0\to k_1,k_1\to k_6,k_2\to k_5,k_4\to k_2,k_5\to k_4,k_6\to k_0,\nonumber\\
&\quad N_0\to N_1,N_1\to N_6,N_2\to N_5,N_4\to N_2,N_5\to N_4,N_6\to N_0.
\label{e6outer}
\end{align}
From the action of the Weyl group, we label the ranks as
\begin{align}
&N_0=N,\quad N_1=N+M_1,\quad N_2=2N+M_2,\quad N_3=3N+M_3,\quad N_4=2N+M_4,\nonumber\\
&N_5=2N+M_5,\quad N_6=N+M_6,
\label{e6relarank}
\end{align}
with the ``overall'' rank $N$ multiplied by the marks.
Using the Weyl reflections \eqref{e6weyl} and the outer automorphism \eqref{e6outer}, we can adopt the same working hypothesis and study examples of duality cascades as in figure \ref{d4dualitycascade}.
Then, similarly, the questions on the finiteness and the uniqueness arise.

To answer these questions, we study the polytope associated to duality cascades again.
As previously, we still have three equivalent descriptions.
For example, we can apply the Weyl reflections \eqref{e6weyl} for the quiver diagram with the relative ranks labeled in \eqref{e6relarank} and write down the inequalities for the ${\cal H}$ description.
Or we can consider various configurations with relative ranks vanishing and return to the original combination of levels by applying the Weyl reflections to determine the vertices for the ${\cal V}$ description.
Out of them the most important and the most non-trivial one is the ${\cal Z}$ description, which is obtained from the ${\cal V}$ description by reading off the vectors of the zonotope as in \eqref{d4zonoq} for the $\widehat D_4$ quiver.
After the whole computations, we find that the vectors are given by
\begin{align}
&{\bm k}_1,\;{\bm k}_2,\;{\bm k}_3,\;{\bm k}_4,\;{\bm k}_5,\;{\bm k}_6,\;{\bm k}_{12},\;{\bm k}_{23},\;{\bm k}_{34},\;{\bm k}_{35},\;{\bm k}_{56},\;{\bm k}_{123},\;{\bm k}_{234},\;{\bm k}_{235},\;{\bm k}_{345},\;{\bm k}_{356},
\nonumber\\
&{\bm k}_{1234},\;{\bm k}_{1235},\;{\bm k}_{2345},\;{\bm k}_{2356},\;{\bm k}_{3456},\;{\bm k}_{12345},\;{\bm k}_{23345},\;{\bm k}_{12356},\;{\bm k}_{23456},\;{\bm k}_{123345},\;{\bm k}_{123456},\;{\bm k}_{233456},\nonumber\\
&{\bm k}_{1223345},\;{\bm k}_{1233456},\;{\bm k}_{2334556},\;{\bm k}_{12233456},\;{\bm k}_{12334556},\;{\bm k}_{122334556},\;{\bm k}_{1223334556},\;{\bm k}_{12233344556},
\label{e6vec}
\end{align}
with
\begin{align}
{\bm k}_{1^{n_1}2^{n_2}3^{n_3}4^{n_4}5^{n_5}6^{n_6}}
=\bigg|\sum_{i=1}^6n_ik_i\bigg|(n_1,n_2,n_3,n_4,n_5,n_6).
\end{align}
These vectors correspond to the positive roots as in the case of the $\widehat A$ and $\widehat D$ quivers.

The discrete translations of duality cascades also work similarly as previously.
For the potentially affine nodes \textcircled{\scriptsize 1} and \textcircled{\scriptsize 6} we can compare with the affine node \textcircled{\scriptsize 0} directly.
We read off not only $\Delta_1{\bm M}$ and $\Delta_6{\bm M}$ but also all others $\Delta_i{\bm M}$ since other variables can come in the potentially affine nodes through the Weyl reflections.
As in \eqref{DMk}, the basis of the discrete translations of duality cascades $\Delta_i{\bm M}$ can be read off from the expressions in \eqref{e6vec} by the rule that the coefficient of ${\bm k}_{\cdots}$ in $\Delta_i{\bm M}$ is the times that $i$ appears in the subscripts of ${\bm k}_{\cdots}$.
For example, we have
\begin{align}
&\Delta_3{\bm M}={\bm k}_3+{\bm k}_{23}+{\bm k}_{34}+{\bm k}_{35}+{\bm k}_{123}+{\bm k}_{234}+{\bm k}_{235}+{\bm k}_{345}+{\bm k}_{356}
\nonumber\\
&+{\bm k}_{1234}+{\bm k}_{1235}+{\bm k}_{2345}+{\bm k}_{2356}+{\bm k}_{3456}\nonumber\\
&+{\bm k}_{12345}+2{\bm k}_{23345}+{\bm k}_{12356}+{\bm k}_{23456}+2{\bm k}_{123345}+{\bm k}_{123456}+2{\bm k}_{233456}\nonumber\\
&+2{\bm k}_{1223345}+2{\bm k}_{1233456}+2{\bm k}_{2334556}+2{\bm k}_{12233456}+2{\bm k}_{12334556}+2{\bm k}_{122334556}\nonumber\\
&+3{\bm k}_{1223334556}+3{\bm k}_{12233344556}.
\label{e6DMk}
\end{align}
For the non-affine nodes, the distances between the opposite facets are shortened since the non-trivial marks do not appear in the coefficients.
For example, the vertices on the two opposite facets $\pm M_3+\cdots=0$ are different by
\begin{align}
&{\bm k}_3+{\bm k}_{23}+{\bm k}_{34}+{\bm k}_{35}+{\bm k}_{123}+{\bm k}_{234}+{\bm k}_{235}+{\bm k}_{345}+{\bm k}_{356}
\nonumber\\
&+{\bm k}_{1234}+{\bm k}_{1235}+{\bm k}_{2345}+{\bm k}_{2356}+{\bm k}_{3456}\nonumber\\
&+{\bm k}_{12345}+{\bm k}_{23345}+{\bm k}_{12356}+{\bm k}_{23456}+{\bm k}_{123345}+{\bm k}_{123456}+{\bm k}_{233456}\nonumber\\
&+{\bm k}_{1223345}+{\bm k}_{1233456}+{\bm k}_{2334556}+{\bm k}_{12233456}+{\bm k}_{12334556}+{\bm k}_{122334556}\nonumber\\
&+{\bm k}_{1223334556}+{\bm k}_{12233344556},
\label{e6opposite}
\end{align}
with non-trivial coefficients missing.
By comparing the discrete translation \eqref{e6DMk} and the distance between the vertices on the opposite facets \eqref{e6opposite}, we find again that the polytope tiles the parameter space of relative ranks but with gaps.
These studies imply that the questions on the finiteness and the uniqueness are answered positively under certain restrictions.

\subsection{$\widehat E_7$ quiver}

\begin{figure}[t!]
\centering\includegraphics[scale=0.5,angle=-90]{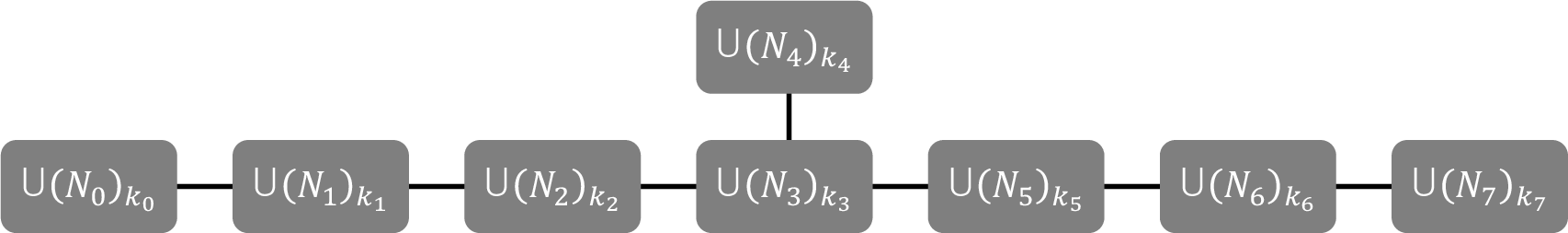}
\caption{$\widehat E_7$ quiver.}
\label{e7}
\end{figure}

The situation for the $\widehat E_7$ quiver is similar.
The duality transformations inspired by the Weyl reflections are given by
\begin{align}
r_1&:k_1\to-k_1,k_0\to k_0+k_1,k_2\to k_2+k_1,N_1\to N_0+N_2-N_1+|k_1|,\nonumber\\
r_2&:k_2\to-k_2,k_1\to k_1+k_2,k_3\to k_3+k_2,N_2\to N_1+N_3-N_2+|k_2|,\nonumber\\
r_3&:k_3\to-k_3,k_2\to k_2+k_3,k_4\to k_4+k_3,k_5\to k_5+k_3,N_3\to N_2+N_4+N_5-N_3+|k_3|,\nonumber\\
r_4&:k_4\to-k_4,k_3\to k_3+k_4,N_4\to N_3-N_4+|k_4|,\nonumber\\
r_5&:k_5\to-k_5,k_3\to k_3+k_5,k_6\to k_6+k_5,N_5\to N_3+N_6-N_5+|k_5|,\nonumber\\
r_6&:k_6\to-k_6,k_5\to k_5+k_6,k_7\to k_7+k_6,N_6\to N_5+N_7-N_6+|k_6|,\nonumber\\
r_7&:k_7\to-k_7,k_6\to k_6+k_7,N_7\to N_6-N_7+|k_7|,
\label{e7weyl}
\end{align}
and the outer automorphism is
\begin{align}
\sigma:k_0\leftrightarrow k_7,k_1\leftrightarrow k_6, k_2\leftrightarrow k_5,N_0\leftrightarrow N_7,N_1\leftrightarrow N_6,N_2\leftrightarrow N_5.
\end{align}
We label the ranks by
\begin{align}
&N_0=N,\quad N_1=2N+M_1,\quad N_2=3N+M_2,\quad N_3=4N+M_3,\quad N_4=2N+M_4,\nonumber\\
&N_5=3N+M_5,\quad N_6=2N+M_6,\quad N_7=N+M_7.
\end{align}

In the ${\cal Z}$ description out of the three equivalent ones, the vectors of the zonotope are given by
\begin{align}
&{\bm k}_{1},{\bm k}_{2},{\bm k}_{3},{\bm k}_{4},{\bm k}_{5},{\bm k}_{6},{\bm k}_{7},{\bm k}_{12},{\bm k}_{23},{\bm k}_{34},{\bm k}_{35},{\bm k}_{56},{\bm k}_{67},{\bm k}_{123},{\bm k}_{234},{\bm k}_{235},{\bm k}_{345},{\bm k}_{356},{\bm k}_{567},\nonumber\\
&{\bm k}_{1234},{\bm k}_{1235},{\bm k}_{2345},{\bm k}_{2356},{\bm k}_{3456},{\bm k}_{3567},{\bm k}_{12345},{\bm k}_{12356},{\bm k}_{23345},{\bm k}_{23456},{\bm k}_{23567},{\bm k}_{34567},\nonumber\\
&{\bm k}_{123345},{\bm k}_{123456},{\bm k}_{123567},{\bm k}_{233456},{\bm k}_{234567},{\bm k}_{1223345},{\bm k}_{1233456},{\bm k}_{1234567},{\bm k}_{2334556},{\bm k}_{2334567},\nonumber\\
&{\bm k}_{12233456},{\bm k}_{12334556},{\bm k}_{12334567},{\bm k}_{23345567},{\bm k}_{122334556},{\bm k}_{122334567},{\bm k}_{123345567},{\bm k}_{233455667},\nonumber\\
&{\bm k}_{1223334556},{\bm k}_{1223345567},{\bm k}_{1233455667},{\bm k}_{12233344556},{\bm k}_{12233345567},{\bm k}_{12233455667},\nonumber\\
&{\bm k}_{122333445567},{\bm k}_{122333455667},{\bm k}_{1223334455667},{\bm k}_{1223334555667},{\bm k}_{12233344555667},\nonumber\\
&{\bm k}_{122333344555667},{\bm k}_{1222333344555667},{\bm k}_{11222333344555667}
\end{align}
which correspond to the positive roots.
We can read off the basis of the discrete translations of duality cascades $\Delta_i{\bm M}$ similarly as in \eqref{DMk} and \eqref{e6DMk}.

\subsection{$\widehat E_8$ quiver}

\begin{figure}[t!]
\centering\includegraphics[scale=0.5,angle=-90]{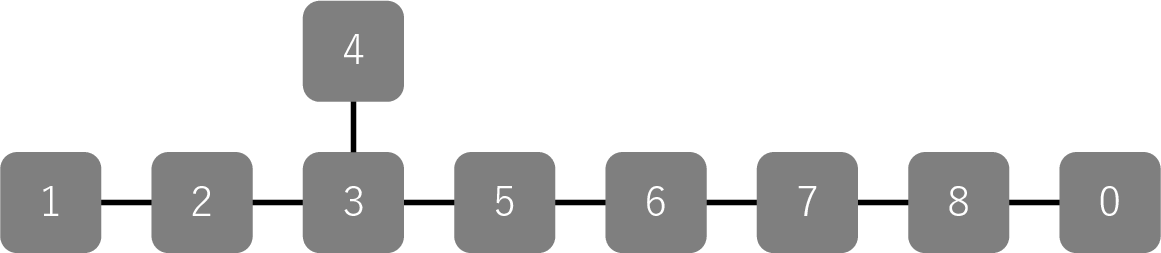}
\caption{$\widehat E_8$ quiver.
Each node $i$ denotes the gauge group factor $\text{U}(N_i)_{k_i}$.}
\label{e8}
\end{figure}

Since there is no automorphism for the $\widehat E_8$ quiver, duality cascades do not work.
Nevertheless, since the Weyl reflection is given similarly, we can still construct the duality transformations and study the polytope associated to duality cascades in the three equivalent descriptions.
Moreover, we can define the discrete translations formally by the same rule as \eqref{DMk} and \eqref{e6DMk}.
We have checked that various geometric properties are given similarly in the group-theoretical language as for the other $\widehat D\widehat E$ quivers.
However, even we define the discrete translations, no facets are moved to the opposite ones without gaps by these translations.

\section{Comments on partition functions}\label{pf}

One of our original motivations for the current study of duality cascades is to understand bilinear relations for the super Chern-Simons theories of the $\widehat D_n$ quivers, such as $\widehat D_4$.
The partition functions of the $\widehat D_n$ quivers were studied in \cite{ADF,MN4} and recently refined in \cite{KN} for $\widehat D_4$ with levels $(-k,0,0,0,k)$ or $q_1=q_2=q_3=q_4=k/2$.
It was unclear why the choice of the levels simplifies the analysis.
Here we propose the reason to be the missing gaps.
In the special choice of the levels, the gaps appearing in tiling the space are filled and this may enable the clean WKB analysis.

In the analysis of $\widehat D_4$, the structure of bilinear relations has not been clarified yet.
As a preliminary step, we would like to look into the geometric picture and propose the ratios of partition functions to be studied under duality cascades.
In the grand partition function $\Xi_{k,M_1,\cdots,M_4}(\kappa)=\sum_{N=0}^\infty\kappa^NZ_{k}(N,N+M_1,2N+M_2,N+M_3,N+M_4)$, some partition functions do not appear due to the non-trivial mark $2$.
After defining the grand partition function, as in \cite{MN6,MN7} for the $\widehat A_3$ quiver, it would be interesting to study the transformations of the partition functions under the discrete translations of duality cascades $Z_k(\Delta_i{\bm M})/Z_k({\bm 0})$ with $\Delta_i{\bm M}$ \eqref{DMq},
\begin{align}
\Delta_1{\bm M}=k(3,4,2,3),\quad
\Delta_3{\bm M}=k(2,4,3,3),\quad
\Delta_4{\bm M}=k(3,6,3,6),
\label{D134M}
\end{align}
and the transformations under other discrete translations involving $\Delta_2{\bm M}=k(4,8,4,6)$ since the corresponding studies in \cite{MN6,MN7} provide hints to bilinear relations for the $\widehat A_3$ quiver.
Note that, although the distance between the corresponding opposite facets is $k(3,6,3,5)$ shortened compared with $\Delta_2{\bm M}$, the facets indeed reduce dimensions for this choice of levels.

\section{Conclusion and discussion}

In this paper we have studied duality cascades for other general affine quivers such as the $\widehat D\widehat E$ quivers.
For the $\widehat A$ quivers, the interpretation through the brane pictures is clear.
By rewriting the working hypothesis of duality cascades in the group-theoretical language, we find the process applicable to more general quivers.
Namely, for the $\widehat D\widehat E$ quivers, although the original pictures of branes are missing, with the duality transformations inspired by the Weyl reflections and the outer automorphisms, we can still follow the process of duality cascades.
It is then natural to ask, as in the $\widehat A$ quivers, whether duality cascades always terminate and whether the endpoint is unique regardless of processes.
Previously in \cite{FMMN,FMS} for the $\widehat A$ quivers, these questions were answered positively by defining the fundamental domain and rephrasing the questions into whether the fundamental domain is a parallelotope.
We try to answer the same questions for the $\widehat D\widehat E$ quivers by the same geometric method.

The main difference is that, unlike the $\widehat A$ quivers where ranks of all nodes can always be compared with that of the affine node and induce duality cascades, here for the $\widehat D\widehat E$ quivers, only some of them with trivial marks can be compared and induce duality cascades.
These nodes are called potentially affine nodes here.
When one of the ranks of these nodes is lower than that of the affine node, we can use the outer automorphism to move it to the affine node.
To bring the levels changed by the outer automorphism back to the original ones, we have to apply a sequence of Weyl reflections.
Then, we find that the relative ranks are transformed by discrete translations.
Since the number of independent discrete translations matches the space dimension, all of these translations are compatible, which means that all the copies fill the space without any overlaps.
However, since some nodes cannot be the affine node through the outer automorphism, if we require their ranks to be larger than that of the affine rank, the copies will have gaps when tiling the space.
This can be seen from the direct computations comparing the discrete translations induced by the potentially affine nodes and the distances between the opposite parallel facets, as provided in the previous sections.
The misalignments come from the missing non-trivial marks in the coefficients of the distances.
Hence, we may attribute the gaps to the non-trivial marks.
We also find that the gaps disappear when we choose the levels appropriately to be those which work well in the previous WKB analysis in \cite{MN4,KN}.
All these studies imply that, with large ranks for the non-affine nodes, duality cascades for the $\widehat D\widehat E$ quivers still teminate in finite steps and the endpoint is uniquely determined by a starting point regardless of processes.
In the following, let us point out some future directions we would like to pursue.

First, most importantly, our duality transformations in \eqref{d4weyl}, \eqref{e6weyl}, \eqref{e7weyl} are based purely on the consistency with the Weyl reflections.
It is critical to understand the dualities from physical gauge theories.
Especially, we are interested in whether the dualities are realized on partition functions, which are localized to matrix models.
Also, it is interesting to understand whether the dualities hold for physical excitations as well.

Second, the polytopes associated to duality cascades for various quivers are mathematically interesting as generalizations from the permutohedrons \cite{JSD}.
Especially, we would like to know whether various properties of these polytopes have any interesting physical implications to duality cascades.
It is also interesting to compare the polytope associated to duality cascades for the $\widehat D_4$ quiver with the four-dimensional parallelotopes \cite{DG}.

Third, one of our original motivations for the current study of duality cascades is to understand bilinear relations \cite{BGT,BGKNT,MN6} for the $\widehat D$ quivers such as $\widehat D_4$ \cite{MN4,KN}.
As we have provided the geometric picture, the next step would be to investigate the ratios of partition functions under discrete translations of duality cascades \eqref{D134M}.
It is also interesting to study various symmetries for the current setup.

Fourth, we have observed that the gaps appearing in tiling are filled with the special choice of levels for $\widehat D_4$ which simplifies the analysis of partition functions.
It is an interesting question to understand what the general solutions of levels are for each quiver when the gaps are filled.
Also, we would like to compare the results with the condition of supersymmetry enhancement \cite{IK}.

Fifth, it was known \cite{NT} that the Coulomb branch of four-dimensional gauge theories of the $\widehat A$ quivers is isomorphic to the Cherkis bow variety  and the similarity to our current study of duality cascades in three-dimensional gauge theories was pointed out.
It is interesting to compare the similarity for other quivers.

\section*{Acknowledgment}

We are grateful to Heng-Yu Chen, Pei-Ming Ho, Yosuke Imamura, Hiroaki Kanno, Naotaka Kubo, Tomoki Nakanishi, Takahiro Nishinaka, Tomoki Nosaka, Hikaru Sasaki and Shintaro Yanagida for valuable discussions and comments.
The work of S.M. is supported by JSPS Grant-in-Aid for Scientific Research (C) \#22K03598.
S.M. would like to thank Yukawa Institute for Theoretical Physics at Kyoto University for warm hospitality.

\appendix

\section{Non-simply-laced quivers}

In the main text, we have concentrated on the simply-laced quivers $\widehat A\widehat D\widehat E$.
Although the vanishing beta functions for unitary groups select out the simply-laced quivers, non-simply-laced quivers work for orthogonal groups and symplectic groups \cite{GHN}.
In this appendix, we study duality cascades for these non-simply-laced quivers.
Since the structure is mostly identical to the simply-laced quivers, we shall be brief in our description.

\begin{figure}[t!]
\centering\includegraphics[scale=0.5,angle=-90]{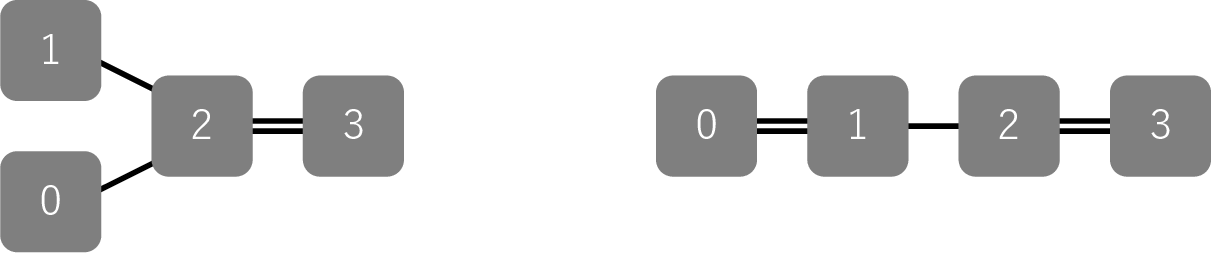}
\caption{Quiver diagrams of type $\widehat B_3$ (left) and type $\widehat C_3$ (right).}
\label{bc3}
\end{figure}

For the non-simply-laced quivers, it is interesting to study the difference between the $\widehat B_n$ and $\widehat C_n$ quivers.
Since the Weyl groups for both quivers are identical, the polytopes should also be identical as zonotopes.
Nevertheless, the $\widehat B_n$ and $\widehat C_n$ quivers are different as the affine Weyl groups and for this reason the discrete translations for duality cascades should be different.
Here we shall concentrate on the case of $\widehat B_3$ and $\widehat C_3$ (see figure \ref{bc3}).
On one hand, the Weyl group for $\widehat B_3$ is
\begin{align}
r_1&:k_1\to-k_1,k_2\to k_2+k_1,N_1\to N_2-N_1+|k_1|,\nonumber\\
r_2&:k_2\to-k_2,k_0\to k_0+k_2,k_1\to k_1+k_2,k_3\to k_3+k_2, N_2\to N_0+N_1+N_3-N_2+|k_2|,\nonumber\\
r_3&:k_3\to-k_3,k_2\to k_2+2k_3,N_3\to 2N_2-N_3+|k_3|,
\label{b3weyl}
\end{align}
and the outer automorphism ${\mathbb Z}_2$ is
\begin{align}
\sigma&:k_0\leftrightarrow k_1,N_0\leftrightarrow N_1.
\end{align}
On the other hand, the Weyl group for $\widehat C_3$ is
\begin{align}
r_1&:k_1\to-k_1,k_0\to k_0+2k_1,k_2\to k_2+k_1,N_1\to 2N_0+N_2-N_1+|k_1|,\nonumber\\
r_2&:k_2\to-k_2,k_1\to k_1+k_2,k_3\to k_3+2k_2,N_2\to N_1+2N_3-N_2+|k_2|,\nonumber\\
r_3&:k_3\to-k_3,k_2\to k_2+k_3,N_3\to N_2-N_3+|k_3|,
\label{c3weyl}
\end{align}
and the outer automorphism ${\mathbb Z}_2$ is
\begin{align}
\sigma&:k_0\leftrightarrow k_3,k_1\leftrightarrow k_2,N_0\leftrightarrow N_3,N_1\leftrightarrow N_2.
\end{align}
Note that the transformations of levels are fixed by the Weyl reflections for simple roots and the transformations of ranks are chosen so that the coefficients of the ``overall'' ranks $N$ in \eqref{NB3} and \eqref{NC3} later are identical to the marks (instead of the comarks).

\begin{figure}[t!]
\centering\includegraphics[scale=0.5,angle=-90]{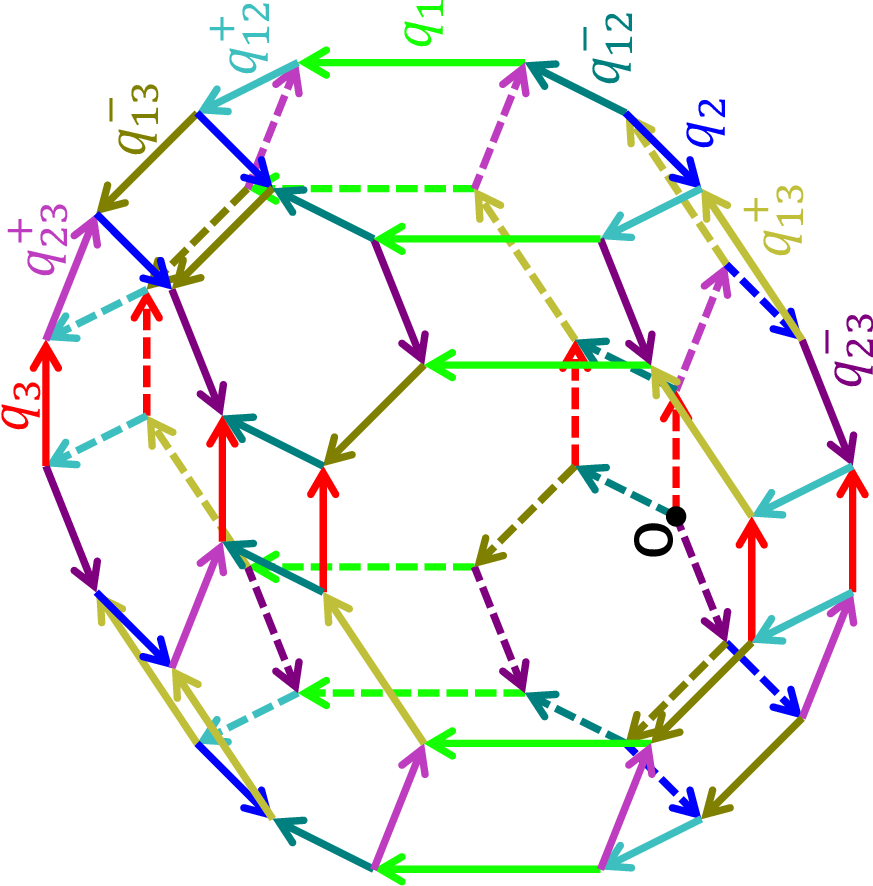}
\caption{The identical polytope associated to duality cascades for the $\widehat B_3$ quiver and the $\widehat C_3$ quiver.}
\label{bc3zono}
\end{figure}

Since, in either of the three descriptions, the polytope associated to duality cascades is constructed from the finite Weyl group and the finite Weyl groups for the two cases are identical, the polytopes are also identical as zonotopes.
Nevertheless, the vectors which generate the zonotopes are different.
If we label the levels for the $\widehat B_3$ quiver as
\begin{align}
k_0=-q_1-q_2,\quad k_1=q_1-q_2,\quad k_2=q_2-q_3,\quad k_3=q_3
\end{align}
the vectors generating the zonotope are
\begin{align}
{\bm q}_1=|q_1|(1,1,1),\quad{\bm q}_{12}^-=|q_{12}^-|(1,0,0),\quad{\bm q}_{12}^+=|q_{12}^+|(1,2,2),\nonumber\\
{\bm q}_2=|q_2|(0,1,1),\quad{\bm q}_{23}^-=|q_{23}^-|(0,1,0),\quad{\bm q}_{23}^+=|q_{23}^+|(0,1,2),\nonumber\\
{\bm q}_3=|q_3|(0,0,1),\quad{\bm q}_{13}^-=|q_{13}^-|(1,1,0),\quad{\bm q}_{13}^+=|q_{13}^+|(1,1,2),
\end{align}
If we label the levels for the $\widehat C_3$ quiver as
\begin{align}
k_0=-2q_1,\quad k_1=q_1-q_2,\quad k_2=q_2-q_3,\quad k_3=2q_3,
\end{align}
the vectors generating the zonotope are
\begin{align}
{\bm q}_1=|q_1|(4,4,2),\quad{\bm q}_{12}^-=|q_{12}^-|(1,0,0),\quad{\bm q}_{12}^+=|q_{12}^+|(1,2,1),\nonumber\\
{\bm q}_2=|q_2|(0,4,2),\quad{\bm q}_{23}^-=|q_{23}^-|(0,1,0),\quad{\bm q}_{23}^+=|q_{23}^+|(0,1,1),\nonumber\\
{\bm q}_3=|q_3|(0,0,2),\quad{\bm q}_{13}^-=|q_{13}^-|(1,1,0),\quad{\bm q}_{13}^+=|q_{13}^+|(1,1,1),
\end{align}
Despite the differences, the zonotopes are given identically in figure \ref{bc3zono}.

\begin{figure}[t!]
\centering
\includegraphics[scale=0.5,angle=-90]{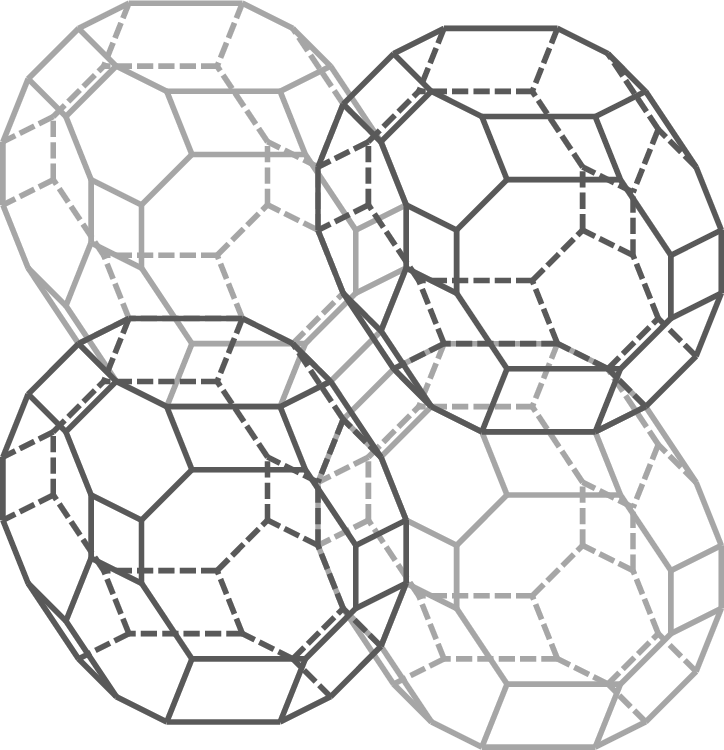}
\qquad\quad\qquad\includegraphics[scale=0.5,angle=-90]{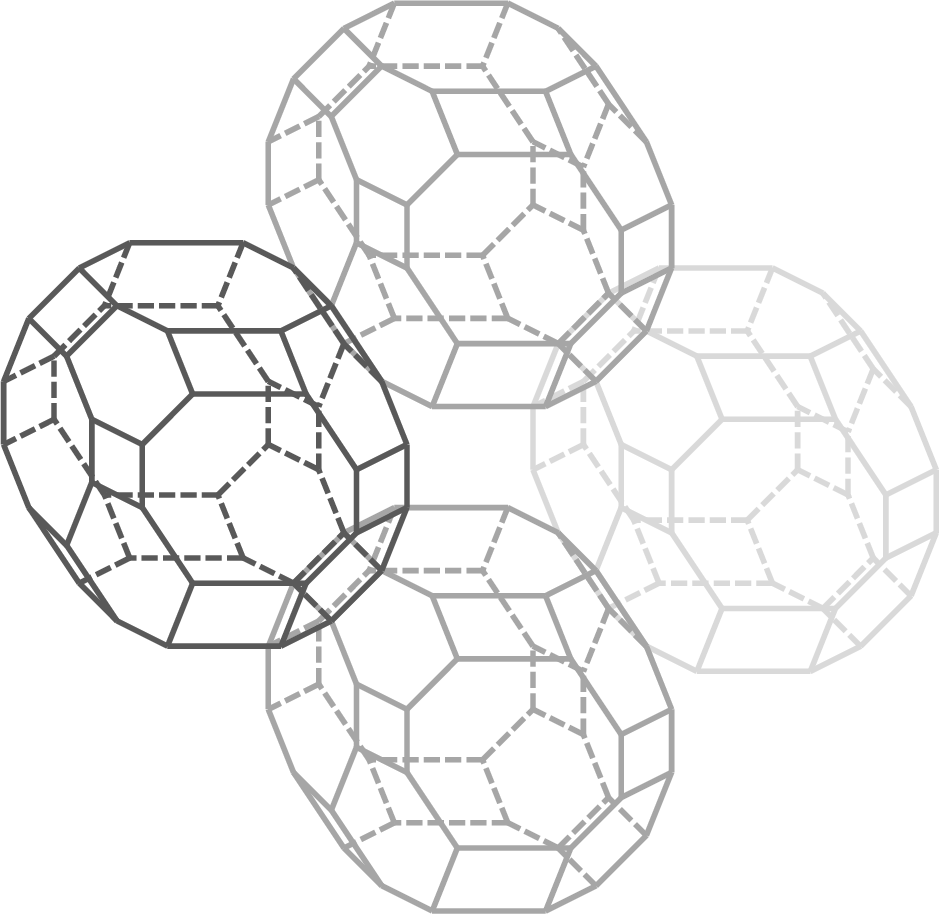}
\caption{Discrete translations for the $\widehat B_3$ quiver (left) and the $\widehat C_3$ quiver (right).
Although the polytopes are identical, the difference in affine quivers causes difference in discrete translations in duality cascades, which leads to different fillings and different gaps in the space.}
\label{bc3filling}
\end{figure}

After fixing the polytopes, we can study duality cascades.
Let us label the ranks for the $\widehat B_3$ quiver by
\begin{align}
N_0=N,\quad N_1=N+M_1,\quad N_2=2N+M_2,\quad N_3=2N+M_3,
\label{NB3}
\end{align}
where the integral coefficients of  the ``overall'' rank $N$ are invariant under the Weyl reflections \eqref{b3weyl}.
When $M_1<0$, we can apply the outer automorphism $\sigma$ and reduce to the original levels using the Weyl reflections.
Then, both the overall rank $N$ and the relative ranks ${\bm M}=(M_1,M_2,M_3)$ are shifted by discrete translations.
We can also apply the Weyl reflections to study the cases where other rank variables $M_i$ appearing in the position of $N_1$ are negative.
Overall, the discrete translations are generated by
\begin{align}
\Delta_1{\bm M}&={\bm q}_1+{\bm q}_{12}^-+{\bm q}_{12}^++{\bm q}_{13}^-+{\bm q}_{13}^+,\nonumber\\
\Delta_2{\bm M}&={\bm q}_1+2{\bm q}_{12}^++{\bm q}_{13}^-+{\bm q}_{13}^++{\bm q}_2+{\bm q}_{23}^-+{\bm q}_{23}^+,\nonumber\\
\Delta_3{\bm M}&={\bm q}_1+2{\bm q}_{12}^++2{\bm q}_{13}^++{\bm q}_2+2{\bm q}_{23}^++{\bm q}_3,
\end{align} 
whose numbers are the same as the space dimensions.
However, the distance between vertices on the two facets $\pm M_2+\cdots=0$ (or $\pm M_3+\cdots=0$) is given by $\Delta_2{\bm M}$ (or $\Delta_3{\bm M}$) with all the non-trivial coefficients $2$ set to $1$.
Similarly, we label the ranks for the $\widehat C_3$ quiver by
\begin{align}
N_0=N,\quad N_1=2N+M_1,\quad N_2=2N+M_2,\quad N_3=N+M_3,
\label{NC3}
\end{align}
where the integral coefficients of  the ``overall'' rank $N$ are invariant under the Weyl reflections \eqref{c3weyl}.
When $M_3<0$, we can apply the outer automorphism $\sigma$ and reduce to the original levels using the Weyl reflections.
Then, both $N$ and ${\bm M}=(M_1,M_2,M_3)$ are shifted by discrete translations.
We also apply the Weyl reflections to study the cases where other rank variables $M_i$ appearing in the position of $N_3$ are negative.
Overall, the discrete translations are generated by
\begin{align}
\Delta_1{\bm M}&=2{\bm q}_1+{\bm q}_{12}^-+{\bm q}_{12}^++{\bm q}_{13}^-+{\bm q}_{13}^+,\nonumber\\
\Delta_2{\bm M}&=2{\bm q}_1+2{\bm q}_{12}^++{\bm q}_{13}^-+{\bm q}_{13}^++2{\bm q}_2+{\bm q}_{23}^-+{\bm q}_{23}^+,\nonumber\\
\Delta_3{\bm M}&={\bm q}_1+{\bm q}_{12}^++{\bm q}_{13}^++{\bm q}_2+{\bm q}_{23}^++{\bm q}_3,
\end{align} 
whose numbers are the same as the space dimensions.
However, the distance between vertices on the two facets $\pm M_1+\cdots=0$ (or $\pm M_2+\cdots=0$) is given by $\Delta_1{\bm M}$ (or $\Delta_2{\bm M}$) with all the non-trivial coefficients $2$ set to $1$.

For these reasons, for both the cases of the $\widehat B_3$ and $\widehat C_3$ quivers, the polytopes do not form parallelotopes and have gaps in space.
However, the gaps for $\widehat B_3$ and those for $\widehat C_3$ are different.
We depict the space tilings in figure \ref{bc3filling}.
Interestingly, these examples show that, even the polytopes are the same, the space tilings can be different depending on quivers.

\end{document}